\documentclass[12pt]{iopart}

\usepackage{cite}
\usepackage{graphicx}
\usepackage{url}
\usepackage{xcolor}
\usepackage{tikz,pgf}
\usetikzlibrary{arrows.meta}
\usepackage{multirow}

\expandafter\let\csname equation*\endcsname=\relax
\expandafter\let\csname endequation*\endcsname=\relax
\usepackage{amsmath,amssymb,mhchem}

\newcommand{\eq}[1]{\begin{equation} #1 \end{equation}}
\newcommand{\elmx}[3]{\langle #1|#2|#3 \rangle}
\newcommand{\ket}[1]{|#1\rangle}

\begin{document}

\title{Octupole deformation in quasiparticle states of odd-mass and
  odd-odd nuclei}

\author{N. Kontowicz$^1$, L. Bonneau$^1$, J. Bartel$^2$,
  H. Molique$^2$, N. Minkov$^3$ and Meng-Hock Koh$^{4}$}

\address{$^1$LP2I Bordeaux, UMR 5797, Université de Bordeaux, CNRS,
  F-33170, Gradignan, France}
\address{$^2$IPHC, UMR 7178, Universit\'e de Strasbourg, CNRS,
  F-67037, Strasbourg, France}
\address{$^3$Institute of Nuclear Research and Nuclear Energy, Bulgarian
  Academy of Sciences, Tzarigrad Road 72, BG-1784 Sofia, Bulgaria}
\address{$^4$Department of Physics, Faculty of Science, Universiti
  Teknologi Malaysia, 81310 Johor Bahru, Johor, Malaysia}

\begin{abstract}

  As a follow up of [Phys. Scr. 99 055305 (2024)], where we studied
  axial octupole shapes in two-quasiparticle states of even-even
  nuclei, we investigate this type of shapes in odd-mass and
  odd-odd well-deformed nuclei, using the Skyrme--Hartree--Fock--BCS
  approach with selfconsistent blocking and a constraint on the
  expectation value $Q_{30}$ of the axial octupole moment operator. To
  interprete the pattern of the resulting deformation energy curve as
  a function of $Q_{30}$, we extend the perturbative mechanism of
  Ref.~\cite{Bonneau24_PhysScr99}. We deduce selection rules which can
  predict, from the single-particle spectra at $Q_{30} = 0$, whether
  in a given multi-quasiparticle state the deformation energy curve
  has a local minimum at a vanishing or a finite value of
  $Q_{30}$. The predictions of this perturbative mechanism are
  compared with actual Skyrme--Hartree--Fock--BCS calculations
  with a constraint on the expectation value $Q_{30}$. Overall
  we obtain a qualitative agreement and we show that quantitative
  predictions are limited by the role of pairing correlations and
  strong octupole coupling between quasi-degenerate members of
  a single-particle parity doublet.

\end{abstract}


\submitto{\PS}

\maketitle

%
%
%
%

\section{Introduction}

In Ref.~\cite{Bonneau24_PhysScr99} hereafter called
paper I, we studied axial octupole deformation in
two-quasiparticle states of even-even nuclei by performing 
Skyrme--Hartree--Fock--BCS (SHFBCS) calculations with selfconsistent blocking
and a constraint on the expectation value of the axial octupole moment
operator $\widehat Q_{30}$, defined by
\eq{
  \label{eq_Q30_def}
  \widehat Q_{30} =
  \sqrt{\frac{7}{16\pi}}\,\big(2z^3-3z(x^2+y^2)\big) \,.
}
We showed that the qualitative features of the obtained deformation
energy curve can be understood within a perturbative mechanism
involving selection rules of $\widehat Q_{30}$ between asymptotic
Nilsson configurations and the shape of the ground-state configuration
at the miminum of the deformation energy curve.

In the present paper, we generalize this perturbative mechanism to
odd-mass and odd-odd nuclei. Then to assess its robustness, we
investigate axial octupole deformation in selected odd-mass and
odd-odd nuclei of the $A \sim 150$ and $A\sim 230$ mass regions. To do
so we compute deformation energy curves within the 
SHFBCS approach with self-consistent blocking as
in paper I and compare the existence of a local minimum at finite
or vanishing octupole deformation with the predictions of the
perturbative mechanism.

%
%
%
%

\section{Theoretical framework}

%
%

\subsection{Selfconsistent mean-field framework and calculational details}

To describe low-lying bandhead states of quasiparticle type in
well-deformed nuclei, we adopt a many-body approach relying on
an energy-density functional (EDF) of the Skyrme type with BCS
treatment of pairing correlations and self-consistent blocking of the
unpaired nucleons. This approach can be viewed as a Hartree--Fock--BCS (HFBCS)
approximation to the nuclear many-body problem with a so-called 
``density-dependent'' effective nucleon-nucleon interaction of the
Skyrme type. The blocking treatment of unpaired nucleons breaks the
time-reversal symmetry at the one-body level because it generates non
vanishing spin and current vector fields. In the mean-field channel
we use the SIII parametrization of the Skyrme EDF~\cite{Beiner75} in
the ``minimal'' scheme as explained in Ref.~\cite{Bonneau15_PRC91},
for which the other time-odd fields, namely the spin-gradient and tensor fields, are neglected.
In the pairing channel we use constant pairing matrix elements parametrized
and determined as in Ref.~\cite{Nor19_PRC99}.

To represent the single-particle states of given isospin index $\tau$ ($\tau
=$ n for neutrons or $\tau =$ p for protons), we use an axially-deformed
harmonic oscillator basis~\cite{Damgaard69,Vautherin73,FQKV}. We
provide some calculational details about its parameters in Appendix~A.
The matrix of the one-body Hartree--Fock Hamiltonian $\widehat h$,
which depends on the single-particle states, is diagonalized in this
basis in an iterative process. The energy density, the multipole
moments of the nucleon density and the matrix elements of $\widehat h$
are calculated by integration over space through quadratures using 30
Gauss--Hermite mesh points in the $z$ direction and 15 Gauss--Laguerre
mesh points in the direction orthogonal to the symmetry
axis~\cite{Vautherin73}. 

The basis parameters $N_0$, $b$ and $q$ (defined in Appendix~A)
used in the present work are:
\begin{itemize}
\item $N_0=14$, $b=0.50$, $q=1.15$ in $^{144}$Ba and $^{145}$Ba nuclei
\item $N_0=14$, $b=0.495$, $q=1.18$ in $^{150}$Sm, 
  $^{154}$Gd and neighboring odd-mass and odd-odd nuclei
\item $N_0=16$, $b=0.48$, $q=1.20$ in all considered nuclei of the $A \sim 230$
  region,
\end{itemize}
and the pairing-strength parameters
$G_0^{(\tau)}$ for nucleons of isospin $\tau$ (defined in, e.g.,
Ref.~\cite{Bonneau15_PRC91}) are
\begin{itemize}
\item $G_0^{(n)} = 16.26$ MeV and $G_0^{(p)} = 15.27$ MeV in the
  $A\sim 150$ nuclei, as determined in Ref.~\cite{Nor19_PRC99}
\item $G_0^{(n)} = 16.0$ MeV and $G_0^{(p)} = 14.4$ MeV in the
  $A\sim 230$ nuclei, as the set 1 of Ref.~\cite{Minkov24_PRC110}.
\end{itemize}

Finally, at each Hartree--Fock iteration, the BCS equations (see, e.g.,
Ref.~\cite{Ring-Schuck80}) are solved also iteratively. In these
equations all single-particle states except the blocked states and
their conjugate states are included with a smearing factor of the same
form as in paper~I and Ref.~\cite{Krieger90_NPA517}: 
\eq{
  f(e_i^{(\tau)}) = \big[1 +
    \exp\big((e_i^{(\tau)} - X - \lambda_{\tau})/\mu\big)\big]^{-1/2}
}
where $e_i^{(\tau)}$ is the energy of the single-particle state
$\ket{\psi_i^{(\tau)}}$ of isospin index $\tau$, $X \!=\! 6\;$MeV is a
cutoff parameter, $\lambda_{\tau}$ is the chemical potential and $\mu
\!=\! 0.2$~MeV is a diffuseness parameter. Note that in paper I, the
power $-1$ in the expression of the smearing factor $f$ is a misprint
and should read $-1/2$.

%
%

\subsection{Deformation energy curves of quasiparticle configurations}

\label{def_energy_curve}

Using the notation of Appendix~A we denote by
$\Omega^{\pi}[Nn_{z}\Lambda]$ the asymptotic Nilsson quantum numbers
of the leading contribution in the expansion of a blocked
single-particle state $\ket{\psi_a^{(\tau)}}$ in the axially-deformed
harmonic-oscillator basis. A multi-quasiparticle configuration
$\mathcal C$ being characterized 
by the blocked single-particle states, we use their dominant Nilsson
quantum numbers to label the configuration $\mathcal C$, which can
thus be written as $\mathcal C =
(\Omega_1^{\pi_1}[N_1n_{z,1}\Lambda_1],\cdots)_n
(\Omega_1^{\prime \, \pi'_1}[N'_1n'_{z,1}\Lambda'_1],\cdots)_p$.

To describe the response of a nucleus in a configuration $\mathcal C$
to the axial octupole deformation operator
$\widehat Q_{30}$ within the nuclear EDF approach, we minimize
the total energy $E_{\mathcal C}(Q_{30})$ of this nucleus in the normalized
many-body state $\ket{\Psi}$ of BCS type (with blocking) with
the constraint that $\elmx{\Psi}{\widehat Q_{30}}{\Psi}$ takes a given
value $Q_{30}$. By varying the value of $Q_{30}$ using a quadratic penalty
function as explained in detail in Ref.~\cite{FQKV}, we generate the
deformation energy curve for the studied deformation mode 
and the desired configuration $\mathcal C$. The local minimum along this
curve thus corresponds, within the mean-field picture, to the
energetically most favorable shape of the considered nucleus in the
configuration~$\mathcal C$.

From a numerical point of view, the octupole-deformation-energy curve
is obtained by a smooth interpolation of a set of tabulated points. To
generate these points, we proceed as follows. First, we
compute the HFBCS solution starting from a Woods--Saxon
potential with $\beta_2 = 0.25$ (all studied nuclei are expected to be
prolate deformed), all other deformation parameters $\beta_{\ell}$ being set
to~0. Because the axially-deformed harmonic-oscillator basis is
made of eigenvectors of the parity operator, the selfconsistent
symmetry theorem guarantees that the single-particle states also have
good parity at each Hartree--Fock iteration until convergence, although
parity symmetry is not enforced. We thus reach the first point of the
octupole-deformation-energy curve with $\elmx{\Psi}{\widehat
  Q_{30}}{\Psi} = 0$. It is worth noting that this solution is the
same as the one obtained by imposing parity symmetry, which we checked
numerically in a few cases.

Then, to get the second point of this curve, we
start the HFBCS iterative process from the converged
Hartree--Fock hamiltonian of the first point, with
the constraint $\elmx{\Psi}{\widehat Q_{30}}{\Psi} = d Q_{30}$ where
the axial-octupole moment step is
$d Q_{30} = \dfrac{3}{4\pi}A^2r_0^3\,d\beta_3$ using $r_0 = 1.2$~fm and
the step $d\beta_3 = 0.02$. As explained in Ref.~\cite{FQKV},
the target value of a quadratic constraint differs from the converged
value because of the slope of the deformation-energy curve. Therefore
the second point of this curve is such that $\elmx{\Psi}{\widehat
  Q_{30}}{\Psi} = dQ_{30}$. All subsequent points of the 
deformation-energy curve are obtained in a similar way by using an
initial Hartree--Fock hamiltonian of the converged solution at the
previous point and by adding $dQ_{30}$ to the constraint
$\elmx{\Psi}{\widehat Q_{30}}{\Psi}$ of the previous point.

Before closing this section we emphasize that only the studied
deformation mode (here the axial octupole deformation) is subject to a
constraint, because we are interested in the lowest-energy solution as
a function of $Q_{30}$.

%
%
%
%
%

\section{Perturbative mechanism of deformation}

At the heart of the mechanism of deformation developed in paper I for
an even-even nucleus are the Koopmans approximation of
$E_{\mathcal C}(Q_{30})$, neglecting pairing, and the perturbative
expansion of the single-particle (s.p.) energies at lowest order in
$Q_{30}$. Here we want to extend this mechanism to any type of nucleus
along the lines of Ref.~\cite{Thesis}.

In an even-even $(N,Z)$ nucleus, the ground state at the Hartree--Fock
approximation corresponds to the lowest-energy $N$ neutron
single-particle states and $Z$ proton states fully occupied. In
contrast, the ground-state configuration of odd-mass or odd-odd nuclei
corresponds to a blocking structure which depends on the considered 
nucleus and is not a priori known. Therefore, for a unifying
description of any kind of nucleus, we need to introduce an even-even 
reference nucleus (core): 
\begin{itemize}
\item the studied nucleus itself if it is even-even;
\item the $(N-1,Z)$ nucleus if the considered nucleus has
  an odd number $N$ of neutrons and even number $Z$ of protons, 
  or the $(N,Z-1)$ nucleus in the opposite case;
\item either of the four neighbouring $(N-1,Z-1)$, $(N-1,Z+1)$,
  $(N+1,Z-1)$ and $(N+1,Z+1)$ even-even nuclei. 
\end{itemize}

%
%

\subsection{Energy of a given configuration at vanishing
  axial-octupole deformation}

First we consider the HFBCS ground-state solution
$\ket{\Psi_0}$ of the even-even $(N,Z)$ reference nucleus at vanishing
axial octupole deformation. From its s.p. states and
spectrum we build the lowest-energy independent-particle (normalized)
state $\ket{\Phi_0}$
\eq{
  \label{ket_Phi0}
  \ket{\Phi_0} =
  \mathrm{n}^{\dagger}_{k_1}\mathrm{n}^{\dagger}_{\overline k_1}
  \cdots
  \mathrm{n}^{\dagger}_{k_{N/2}} \mathrm{n}^{\dagger}_{\overline{k}_{N/2}}
  \mathrm{p}^{\dagger}_{l_1} \mathrm{p}^{\dagger}_{\overline l_1}
  \cdots
  \mathrm{p}^{\dagger}_{p_{Z/2}} \mathrm{p}^{\dagger}_{\overline{l}_{Z/2}}
  \ket 0
}
where $\ket 0$ is the particle-vacuum normalized state
and $\mathrm{n}^{\dagger}_x$, $\mathrm{n}_x$ are, respectively,
creation and annihilation operators of the single-neutron state
$\ket{\psi_x^{(\rm n)}}$. Similarly $\mathrm{p}_y^{\dagger}$,
$\mathrm{p}_y$ denote proton creation and annihilation operators of
the single-proton state $\ket{\psi_y^{(p)}}$. The s.p.
state $\ket{\psi_x^{(\tau)}}$ of isospin index $\tau$ is an eigenstate
of the Hartree--Fock Hamiltonian $\hat h$ of the selfconsistent
solution $\ket{\Phi_0}$ with eigenvalue $e_x^{(\tau)}$:
\eq{
  \widehat h \ket{\psi_x^{(\tau)}} = e_x^{(\tau)}
  \ket{\psi_x^{(\tau)}} \,,
}
and $\ket{\psi_{\overline x}^{(\tau)}} = \widehat{\mathcal T}
\ket{\psi_x^{(\tau)}}$ is its time-reversed state. Because the
time-reversal operator $\widehat{\mathcal T}$ commutes with $\widehat
h$ in the ground-state configuration of an even-even nucleus, we have
the Kramers degeneracy $e_{\overline x}^{(\tau)} = e_{x}^{(\tau)}$.
In Eq.~(\ref{ket_Phi0}) the s.p. states occupied in
$\ket{\Phi_0}$ are ordered by increasing value of $e_x^{(\tau)}$.

Then we use the Koopmans approximation to describe the
HFBCS solution $\ket{\Psi}$ at vanishing octupole
deformation corresponding to the configuration $\mathcal C$. From the
s.p states $\ket{\psi_x^{(\tau)}}$ associated with the
HFBCS solution of the even-even reference nucleus with
$Q_{30} = 0$, we build the independent-particle state $\ket{\Phi}$ as
a multi-particle--multi-hole (mp-mh) configuration with respect to the
reference state $\ket{\Phi_0}$ 
\eq{
  \label{Phi_vs_Phi0}
  \ket{\Phi} =
  \mathrm{n}^{\dagger}_{\alpha_1} \cdots \mathrm{n}^{\dagger}_{\alpha_n} 
  \mathrm{n}_{i_m} \cdots \mathrm{n}_{i_1} 
  \mathrm{p}^{\dagger}_{\beta_1} \cdots \mathrm{p}^{\dagger}_{\beta_p} 
  \mathrm{p}_{j_q}\cdots \mathrm{p}_{j_1} \ket{\Phi_0} \,.
}
The value $E_{\mathcal C}[\Phi]$ taken by the nuclear EDF for the mp-mh configuration
$\ket{\Phi}$ is given by
\eq{
  \label{E_Phi_Q30=0}
  E_{\mathcal C}[\Phi] \approx E_{\rm GS}[\Phi_0] +
  \sum_{x=1}^n e^{(\rm n)}_{\alpha_x} - \sum_{x=1}^m
  e^{(\rm n)}_{i_x} + \sum_{x=1}^p e^{(\rm p)}_{\beta_x} - \sum_{x=1}^q
  e^{(\rm p)}_{j_x} \,,
}
where $E_{\rm GS}[\Phi_0]$ is the total energy of the even-even
reference nucleus in the independent-particle state $\ket{\Phi_0}$
corresponding to the ground-state configuration. 

%
%

\subsection{Energy of the configuration at small axial-octupole
  deformation}

From the HFBCS solution $\ket{\Psi'_0}$ of the even-even
reference nucleus in its ground-state configuration with
$\elmx{\Psi'_0}{\widehat Q_{30}}{\Psi'_0} = Q_{30} \ne 0$ and the
HFBCS solution $\ket{\Psi'}$ of the studied
nucleus in the configuration $\mathcal C$ with
$\elmx{\Psi'}{\widehat Q_{30}}{\Psi'} = Q_{30}$, we build in the same
way as in the previous subsection the independent-particle states
$\ket{\Phi'_0}$ and $\ket{\Phi'}$. If $Q_{30}$ is small enough, we
can still write $\ket{\Phi'}$ as a mp-mh
configuration with respect to the reference state $\ket{\Phi'_0}$ of
the same form as in the relation (\ref{Phi_vs_Phi0}) between
$\ket{\Phi}$ and $\ket{\Phi_0}$:
\eq{
  \label{Phi'_vs_Phi'0}
  \ket{\Phi'} =
  \mathrm{n}^{\dagger}_{\alpha'_1} \cdots \mathrm{n}^{\dagger}_{\alpha'_n} 
  \mathrm{n}_{i'_m} \cdots \mathrm{n}_{i'_1} 
  \mathrm{p}^{\dagger}_{\beta'_1} \cdots \mathrm{p}^{\dagger}_{\beta'_p} 
  \mathrm{p}_{j'_q}\cdots \mathrm{p}_{j'_1} \ket{\Phi'_0} \,,
}
where $\mathrm{n}^{\dagger}_{x'}$, $\mathrm{n}_{x'}$ are,
respectively, creation and annihilation
operators of the single-neutron state $\ket{\psi_x^{\prime\,(\rm n)}}$. Similarly
$\mathrm{p}_{y'}^{\dagger}$, $\mathrm{p}_{y'}$ denote proton creation and annihilation
operators of the single-proton state $\ket{\psi_y^{\prime \,(p)}}$. The
single-particle state $\ket{\psi_x^{\prime \, (\tau)}}$ is an
eigenstate of the Hartree--Fock Hamiltonian $\hat h'$ of the 
selfconsistent solution $\ket{\Psi'_0}$ with eigenvalue
$e_x^{\prime \,(\tau)}$:
\eq{
  \widehat h' \ket{\psi_x^{\prime \,(\tau)}} = e_x^{\prime \, (\tau)}
  \ket{\psi_x^{\prime \, (\tau)}} \,.
}
Similarly to Eq.~(\ref{E_Phi_Q30=0}) we can thus write 
\eq{
  \label{E_Phi'_Q30}
  E_{\mathcal C}[\Phi'] \approx E_{\rm GS}[\Phi'_0] +
  \sum_{x=1}^n e^{\prime \, (\rm n)}_{\alpha_x} - \sum_{x=1}^m
  e^{\prime \, (\rm n)}_{i_x} + \sum_{x=1}^p e^{\prime \, (\rm p)}_{\beta_x} - \sum_{x=1}^q
  e^{\prime \, (\rm p)}_{j_x} 
}

Then we approximate the relation between $\widehat h'$ and
$\widehat h$ by
\eq{
  \widehat h' \approx \widehat h + \widehat u
}
where $\widehat u$ is a local spin-independent one-body
potential proportional to $\widehat Q_{30}$. We have checked that
adding a spin-orbit contribution to $\widehat u$ does not alter the
discussion of results in section~\ref{sec_res}. This perturbing potential 
commutes with $\widehat{\mathcal T}$ and the projection
$\widehat J_z$ on the symmetry axis of the total angular momentum
operator of a nucleon. It is worth recalling that $\widehat{\mathcal T}$ and 
$\widehat J_z$ anticommute. At lowest non-vanishing order in $\widehat u$,
the eigenstates and eigenvalues of $\widehat h'$ are given by   
  \begin{align}
    \ket{\psi^{\prime(\tau)}_x} & \approx \ket{\psi_x^{(\tau)}}
      + \sum_{y\ne x}
      \frac{\elmx{\psi_y^{(\tau)}}{\widehat u}{\psi_x^{(\tau)}}}
           {e_y^{(\tau)}-e_x^{(\tau)}} \,
      \ket{\psi_y^{(\tau)}} \\
      e^{\prime(\tau)}_x & \approx e^{(\tau)}_x - \sum_{y \ne x}
      \frac{|\elmx{\psi_y^{(\tau)}}{\widehat u}{\psi_x^{(\tau)}}|^2}
           {e_y^{(\tau)} - e_x^{(\tau)}} \,.
  \end{align}

  Finally the perturbed energy $E_{\mathcal C}[\Phi']$ can be
  approximately written in the form
\eq{
  \label{E_Phi'_perturb}
  E_{\mathcal C}[\Phi'] \approx E_{\mathcal C}[\Phi] + \Delta E_0 + \Delta e
}
where we have set
\begin{align}
  \label{Delta_E0}
  \Delta E_0 & = E_{\rm GS}[\Phi'_0] - E_{\rm GS}[\Phi_0] \\
  \label{Delta_e}
  \Delta e & = \Delta e^{(\rm n)} + \Delta e^{(\rm p)}
\end{align}
with
\begin{align}
  \label{Delta_en}
  \Delta e^{(\rm n)} & = \sum_{x=1}^n \Big(e^{\prime (\rm n)}_{\alpha_x}
  -e^{(\rm n)}_{\alpha_x}\Big) - \sum_{x=1}^m
  \Big(e^{\prime (\rm n)}_{i_x}-e^{(\rm n)}_{i_x}\Big) \nonumber \\
  & = \sum_{x=1}^n \sum_{y\notin \{\alpha_1\cdots\alpha_n\}}
  \frac{|\elmx{\psi_y^{(\rm n)}}{\widehat u}{\psi_{\alpha_x}^{(\rm n)}}|^2}
       {e_{\alpha_x}^{(\rm n)} - e_y^{(\rm n)}}
  - \sum_{x=1}^m \sum_{y\notin \{i_1\cdots i_m\}}
  \frac{|\elmx{\psi_y^{(\rm n)}}{\widehat u}{\psi_{i_x}^{(\rm n)}}|^2}
       {e_{i_x}^{(\rm n)} - e_y^{(\rm n)}}
\end{align}
and
\begin{align}
  \label{Delta_ep}
  \Delta e^{(\rm p)} & = 
  \sum_{x=1}^p \Big(e^{\prime (\rm p)}_{\beta_x}-e^{(\rm n)}_{\beta_x}\Big)
  - \sum_{x=1}^q \Big(e^{\prime (\rm p)}_{j_x}-e^{(\rm p)}_{j_x}\Big)
  \nonumber \\
  & \phantom{=}
  \sum_{x=1}^p \sum_{y\notin \{\beta_1\cdots\beta_p\}}
  \frac{|\elmx{\psi_y^{(\rm p)}}{\widehat u}{\psi_{\beta_x}^{(\rm p)}}|^2}
       {e_{\beta_x}^{(\rm p)} - e_y^{(\rm p)}}
  - \sum_{x=1}^q \sum_{y\notin \{j_1\cdots j_p\}}
  \frac{|\elmx{\psi_y^{(\rm p)}}{\widehat u}{\psi_{j_x}^{(\rm p)}}|^2}
       {e_{j_x}^{(\rm p)} - e_y^{(\rm p)}} \,.  
\end{align}
The quantity $\Delta E_0$ approximates the actual variation of the total
Hartree--Fock--BCS energy of the even-even reference nucleus in its
ground-state configuration when this nucleus is forced to acquire a
small axial-octupole deformation
\eq{
  \Delta E_0 \approx E_{\rm GS}[\Psi'_0] - E_{\rm GS}[\Psi_0] \,.
}
The perturbative correction $\Delta e$ represents an approximation of
the effect of an axial-octupole perturbation on the excitation energy of
the studied nucleus in its configuration~$\mathcal C$ with respect to
the ground state with imposed $Q_{30} = 0$ of the even-even reference
nucleus. We call $\Delta e$ the excitation-energy correction.

%
%

\subsection{The four scenarios of octupole deformation}

\label{subsec_scenarios}

\begin{figure}[b]
  \centering
  \rotatebox{90}{
    \includegraphics[width=0.85\textheight]{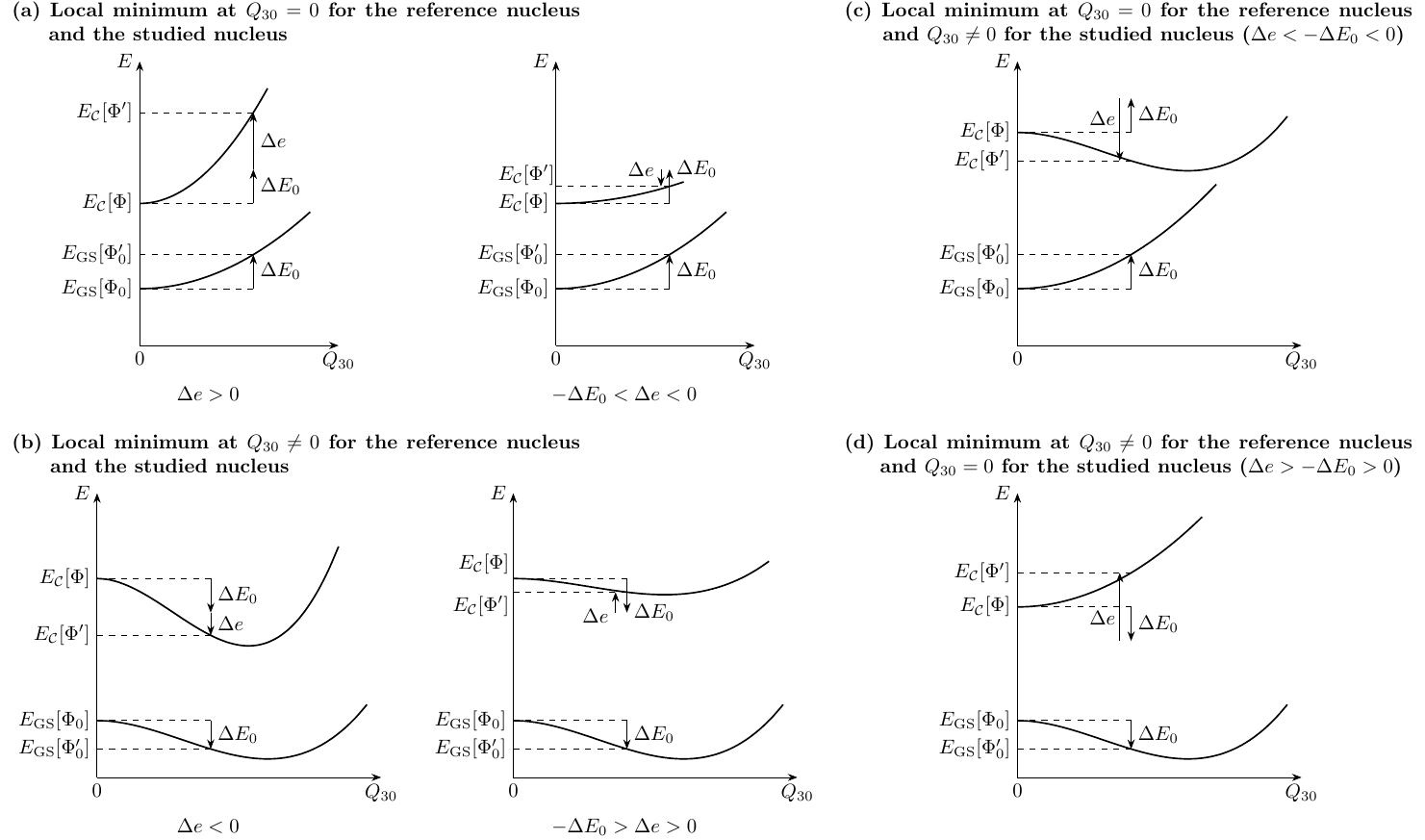}
    }
  \caption{Schematic representation of the four scenarios for the
    deformation energy curves of the studied nucleus and the even-even
    reference nucleus.
    \label{fig_scenarios}}
\end{figure}

From Eqs.~(\ref{E_Phi'_perturb}) to (\ref{Delta_e}) we deduce that four
scenarios are expected depending on the signs of $\Delta E_0$ and
$\Delta E_0+\Delta e$. These scenarios are represented in a
schematic way in Fig.~\ref{fig_scenarios}.

In the first two scenarios, the studied nucleus in its configuration
$\mathcal C$ has the same type of shape as the even-even reference
nucleus in its ground state:
\begin{itemize}
  \item Scenario (a): the studied deformation is not energetically
    favorable in the ground-state configuration of the even-even
    reference nucleus ($\Delta E_0 > 0$) and in the configuration
    $\mathcal C$ of the studied nucleus. This happens when the
    correction $\Delta e$ induced by the perturbing potential
    $\widehat u$ is such that $\Delta e > 0$ (left panel of scenario
    (a) in figure~\ref{fig_scenarios}) or $-\Delta E_0 < \Delta e < 0$
    (right panel of scenario (a) in figure~\ref{fig_scenarios}). The
    considered nucleus is thus expected to have a left-right symmetric
    shape ($Q_{30} = 0$) in its configuration~$\mathcal C$.
    In the limiting case where $\Delta e < 0$ almost cancels $\Delta
    E_0$, the nucleus is ``soft'' with respect to octupole deformation.
  \item Scenario (b): the studied deformation is energetically
    favorable in the ground-state configuration of the even-even
    reference nucleus ($\Delta E_0 < 0$) and in the configuration
    $\mathcal C$ of the studied nucleus. This occurs when the
    correction $\Delta e$ is such that $\Delta e < 0$ (left panel of
    scenario (b) in figure~\ref{fig_scenarios}) or $-\Delta E_0 >
    \Delta e > 0$ (right panel of scenario (b) in
    figure~\ref{fig_scenarios}). The considered nucleus is thus
    expected to have an octupole shape ($Q_{30} \ne 0$) in its
    configuration~$\mathcal C$. In the limiting case where $\Delta e >
    0$ almost cancels $\Delta E_0$, the local minimum of the
    corresponding deformation energy curve is very shallow and the
    nucleus is very soft with respect to octupole deformation.
\end{itemize}
In the other two scenarios, axial-octupole deformation occurs either in
the studied nucleus in its configuration $\mathcal C$ or in the 
even-even reference nucleus in its ground state :
\begin{itemize}
  \item Scenario (c): the studied deformation is not energetically
    favorable in the ground-state configuration of the even-even
    reference nucleus ($\Delta E_0 > 0$) but is favorable in the
    configuration $\mathcal C$ of the studied nucleus. This occurs
    when the correction $\Delta e$ is negative and large enough
    in absolute value to counter-balance $\Delta E_0$. The considered
    nucleus is expected to have an octupole shape ($Q_{30} \ne 0$) in
    its configuration $\mathcal C$.
  \item Scenario (d): this is the opposite to scenario (c) and it
    happens when $\Delta e > 0$ is large enough to counter-balance
    $\Delta E_0 < 0$.
\end{itemize}

%
%

\subsection{Quantitative estimation of single-particle energy
  corrections}

\label{subsec_WS}

As explained in subsection \ref{def_energy_curve}, the full calculation of
the Hartree--Fock--BCS solution at the second point of the
deformation-energy curve uses an initial Hartree--Fock hamiltonian
corresponding to the converged solution at the first point, for which
parity is a selfconsistent symmetry. When convergence is reached, we
thus obtain $\elmx{\Psi}{\widehat Q_{30}}{\Psi} \approx d Q_{30}$,
hence $\beta_3 \approx d\beta_3$ (chosen to be 0.02 in the present
work). 

Our goal is to estimate the single-particle energy corrections $\Delta
e_x^{(\tau)}$ solely from a HFBCS calculation without performing the full reflection-asymmetric calculation.
 To do so, we calculate the matrix elements of
$\widehat u$ in Eqs.~(\ref{Delta_en}) and (\ref{Delta_ep}), between
single-particle states of this left-right symmetric solution, assuming
that the converged Hartree--Fock Hamiltonian $\widehat h'$ at the
second point of the octupole-deformation-energy curve can be
approximated by 
\eq{
  \widehat h' = \widehat h + \widehat u_{\rm WS}
}
where $\widehat h$ is the converged Hartree--Fock--Hamiltonian of the
left-right symmetric solution (first point, at $\beta_3 = 0$) and
$\widehat u_{\rm WS}$ is the parity-symmetry breaking part of a
spin-independent local potential $\widehat V_{\rm WS}$ of the
Woods--Saxon form. Its representation in spherical coordinates
$V_{\rm WS}(r,\theta)$ is written as
\eq{
  V_{\rm WS}(r,\theta) = -\dfrac{V_0}{1+e^{(r-R(\theta))/a}}
}
with
\eq{
  R(\theta) = R_0\,\Big[1+\beta_2\,Y_{20}(\theta) + \beta_3 \,
  Y_{30}(\theta) + \beta_4\,Y_{40}(\theta)\Big] \,.
}
To determine the expression of $u_{\rm WS}(r,\theta)$, we expand
$V_{\rm WS}(r,\theta)$ at first order in $\beta_3$:
\eq{
  \label{eq_u_WS}
  u_{\rm WS}(r,\theta) \approx -\beta_3 \, V_0\, \dfrac{R_0}{a} \,
    \Big(\dfrac{V_{\rm WS}^{(0)}(r,\theta)}{V_0}\Big)^2 \,
    e^{(r-R^{(0)}(\theta))/a}\,\dfrac{Q_{30}(r,\theta)}{r^3} \,
    + o(\beta_3)\,,
}
where $Q_{30}(r,\theta)/r^3$ is proportional to $P_3(\cos
  \theta) = (5\cos^3\theta - 3\cos \theta)/2$ (Legendre polynomial of
  degree 3): 
\eq{
\dfrac{Q_{30}(r,\theta)}{r^3} = \sqrt{\dfrac{7}{4\pi}} \,
P_{3}(\cos\theta)
\,,
}
and $V_{\rm WS}^{(0)}(r,\theta)$ and $R^{(0)}(\theta)$ are deduced from $V_{\rm
  WS}(r,\theta)$ and $R(\theta)$ by setting $\beta_3 = 0$.

For numerical applications, the depth $V_0$ and diffuseness $a$
parameters are chosen to be $V_0 =
-50\times\Big(1+0.8\,\dfrac{N-Z}{A}\Big)$~MeV and $a = 0.5$~fm,
and $R_0$ is taken to be the empirical radius of a nucleus with $A$
nucleons, namely $R_0 = r_0 A^{1/3}$ with $r_0 = 1.2$~fm. 
The Bohr axial deformation parameters $\beta_2$ and
$\beta_4$ are estimated from the expectation
values of the quadrupole and hexadecapole moments of the left-right
symmetric solution, respectively:
\begin{align}
  \beta_2 & =
\sqrt{\dfrac{\pi}5} \, \dfrac{\elmx{\Psi}{\widehat Q_{20}}{\Psi}}{\elmx{\Psi}{\widehat
    R^2}{\Psi}} \\
\beta_4 & = 4\pi\, \dfrac{\elmx{\Psi}{\widehat Q_{40}}{\Psi}}{3 A
  R_0^4}
\end{align}
where $\widehat R$ is the local one-body operator associated with the
radial coordinate $r$. Finally the axial octupole deformation parameter is
taken to be $\beta_3 = 0.02$.

%
%

\subsection{Selection rules governing the sign of the
  excitation-energy correction}

\label{subsec_selection_rules}

According to the expression of the excitation-energy correction
$\Delta e$ in Eq.~(\ref{Delta_e}), the sign of $\Delta e$ results from
a competition between various terms in Eq.~\eqref{Delta_en} and Eq.~\eqref{Delta_ep}. Those giving negative
contributions to $\Delta e$ correspond to
\begin{itemize}
\item coupling through $\widehat u$ of blocked states
  $\ket{\psi_\alpha^{(\tau)}}$ unoccupied in $\ket{\Phi_0}$ with other
  unoccupied states above $e_{\alpha}^{(\tau)}$ in the
  single-particle spectrum of isospin index $\tau$;
\item coupling of blocked states $\ket{\psi_i^{(\tau)}}$ occupied
  in $\ket{\Phi_0}$ with other occupied states lying
  below~$e_i^{(\tau)}$.
\end{itemize}
In contrast, terms giving positive contributions to $\Delta e$
are those for which
\begin{itemize}
\item blocked states $\ket{\psi_\alpha^{(\tau)}}$ unoccupied in
  $\ket{\Phi_0}$ are coupled with occupied states
  below~$e_{\alpha}^{(\tau)}$;
\item blocked states $\ket{\psi_i^{(\tau)}}$ occupied in
  $\ket{\Phi_0}$ are coupled with unoccupied states lying
  above~$e_i^{(\tau)}$. 
\end{itemize}
In practice, the low-lying multiquasiparticle configurations
$\mathcal C$ involve blocked single-particle states near the Fermi
level of the even-even reference nucleus. Therefore $\Delta e$ is
particularly sensitive to axial-octupole couplings between these
blocked states and the paired single-particle states closest in
energy. 

To establish simple rules to guess the sign of $\Delta e$ from
inspection of the single-particle spectrum of the left-right symmetric
Hartree--Fock--BCS solution, the perturbing potential $\widehat u$ is
considered to be simply proportional to $\widehat Q_{30}$ as an
approximation of $u_{\rm WS}(r,\theta)$ in Eq.~(\ref{eq_u_WS}).
As shown in the Appendix~B, the selection rules driving the coupling
through the $\widehat Q_{30}$ operator between axially-deformed
harmonic-oscillator basis states $\ket{Nn_z\Lambda\Sigma}$ and
$\ket{N'n'_z\Lambda'\Sigma'}$ are $|N'-N| = 1$ or 3, $|n'_z-n_z| = 1$ or 3,
$\Lambda' = \Lambda$ and $\Sigma' = \Sigma$ (the spin dependence in the
perturbing potential $\widehat u$ is neglected). When applied to
the leading Nilsson configuration of single-particle states
$\ket{\psi_x^{(\tau)}}$ and $\ket{\psi_y^{(\tau)}}$, these selection
rules provide the main mechanism of axial-octupole coupling between
$\ket{\psi_x^{(\tau)}}$ and $\ket{\psi_y^{(\tau)}}$. Subleading Nilsson
configurations are expected to contribute through these selection
rules to a lesser extent.

%
%
%
%
%

\section{Selection of nuclei and configurations}

\label{sec_selection}

In order to test the octupole deformation mechanism presented in the
previous section, we consider nuclei near the known numbers of
neutrons and protons for which octupole correlations or static
deformation is known to occur, namely $N=88$, 138 and $Z=56$, 88 \cite{Butler96}. Then
we select several low-lying configurations expected to illustrate the
various scenarios of Fig.~\ref{fig_scenarios} in odd-$N$ even-$Z$
nuclei, even-$N$ odd-$Z$ nuclei as well as in odd-odd
nuclei. These configurations involve the single-particle orbitals
found to be more or less sensitive to an octupole perturbation.

To identify these orbitals and to perform a first test of the
perturbative mechanism, we proceed in two steps: first an estimation of
single-particle energy corrections from a left-right symmetric
Hartree--Fock--BCS solution to identify some relevant reference
even-even nuclei, then targeted calculations of
octupole-deformation-energy curves in these even-even nuclei to check
the estimates of the first step and identify relevant configurations
and odd-mass and odd-odd nuclei.

\subsection{Estimation of single-particle energy corrections from
  parity-conserved calculations}

\label{subsec_estimation_Delta_e_WS}

First we compute the Hartree--Fock--BCS solutions for the ground-state
configuration of the \ce{^{144}_{56}Ba_{88}} and
\ce{^{228}_{88}Ra_{140}} nuclei with conserved parity symmetry. This
allows to cover the relevant parts of the neutron and proton 
single-particle spectra. From these left-right symmetric solutions,
using the approach of subsection~\ref{subsec_WS}, we compute the
estimates of energy corrections $\Delta e^{(\tau)}_x$ of
single-particle states from the Woods--Saxon perturbing potential. We
obtain the results displayed in
table~\ref{tab_Q30_sensitive_orbitals}.
\begin{table}[h]
  \centering
  \begin{tabular}{cccccccc}
    \hline
    \multirow{2}{*}{Nucleus} &
    \multicolumn{3}{c}{Neutrons} &&
    \multicolumn{3}{c}{Protons} \\
    \cline{2-4} \cline{6-8} 
    & $\Omega^{\pi} [Nn_z\Lambda]$ & $\Delta e^{(\rm n)}$ (keV)
    & $N_{\rm occ.}$ && $\Omega^{\pi} [Nn_z\Lambda]$ &
    $\Delta e^{(\rm p)}$ (keV) & $Z_{\rm occ.}$ \\
    \hline
    \multirow{6}{*}{\ce{^{144}_{56}Ba_{88}}}
    & $5/2^- [523]$ & $-66$ & 96  && $5/2^- [532]$ & $-40$ & 68 \\
    & $3/2^+ [651]$ & $+126$ & 94 && $3/2^+ [411]$ & $+260$ & 66 \\
    & $3/2^- [521]$ & $-114$ & 92 && $7/2^+ [404]$ & $-18$ & 64 \\
    & $1/2^+ [640]$ & $+27$  & 90 && $3/2^- [541]$ & $-265$ & 62 \\
    & $3/2^- [532]$ & $-27$  & 88 && $1/2^- [530]$ & $+46$ & 60 \\
    & $1/2^- [541]$ & $-35$  & 86 && $1/2^+ [420]$ & $-71$ & 56 \\
    \hline
    \multirow{5}{*}{\ce{^{228}_{88}Ra_{140}}}
    & $7/2^- [743]$ & $-71$ & 144 && $5/2^-[523]$ & $-8$ & 94 \\
    & $5/2^- [752]$ & $+51$ & 142 && $3/2^+[631]$ & $-67$ & 92 \\
    & $3/2^+ [631]$ & $+120$ & 140 && $1/2^-[530]$ & $+524$ & 90 \\
    & $5/2^+ [633]$ & $-108$ & 138 && $1/2^+[640]$ & $-526$ & 88 \\
    & $3/2^- [741]$ & $-115$ & 136 && $1/2^+[600]$ & $-40$ & 86 \\
    \hline
  \end{tabular}
  \caption{\label{tab_Q30_sensitive_orbitals}Single-particle levels
    near the neutron and proton Fermi levels of
    \ce{^{144}_{56}Ba_{88}} and \ce{^{228}_{88}Ra_{140}}.
    The third and sixth columns give the energy corrections defined in
    Eqs.~(\ref{Delta_en}) and (\ref{Delta_ep}) and resulting from the
    octupole-deformed perturbing 
  potential $\widehat u_{\rm WS}$ of Eq.~(\ref{eq_u_WS}).
  The fourth and seventh columns give the
  number of neutrons $N_{\rm occ.}$ and protons $Z_{\rm occ.}$ as
  defined in the text.} 
\end{table}
The asymptotic Nilsson quantum numbers provided in this table
correspond to the dominant cylindrical harmonic-oscillator
contribution in single-particle states at $\beta_3 = 0$.
Moreover we report the number of neutrons $N_{\rm occ.}$ and protons
$Z_{\rm occ.}$, respectively, corresponding to a Hartree--Fock filling
of all orbitals from the lowest-energy one up to and included the
considered orbital (taking into account the Kramers degeneracy)
according to the single-particle spectrum of the nucleus in column 1
obtained at $Q_{30} = 0$. At this stage the values obtained for the
$\Delta e^{(\tau)}_x$ energy corrections are only used as a guide to
anticipate the variation of single-particle energies with $\beta_3$
for small octupole deformations.

Some reference even-even nuclei of interest in the rare-earth region
near $A = 150$ are thus the following ones:
\begin{itemize}
\item \ce{^{144}_{56}Ba_{88}} because the single-particle states corresponding to
  $N_{\rm occ.} = 88$ and $Z_{\rm occ.} = 56$ are
  octupole-deformation-driving orbitals (the $N=88$ and $Z=56$ shell
  gaps increase as functions of $\beta_3$);
\item \ce{^{150}_{62}Sm_{88}} in order to study the weakly
  octupole-sensitive $7/2^+[404]$ and $5/2^-[532]$ proton orbitals and 
  the strongly up-sloping $3/2^+[651]$ orbital as low-lying particle
  states just above $Z_{\rm occ.} = 62$, while keeping $N = 88$ as in
  $^{144}$Ba. Moreover this allows to study the octupole-sensitive
  $1/2^-[530]$ proton orbital as a hole state, in contrast with
  $^{144}$Ba in which it lies above the Fermi level;
\item \ce{^{154}_{64}Gd_{90}} to test the mechanism in higher
  level-density sectors of the neutron and proton spectra.
\end{itemize}

In the actinide region near $A = 230$ the $3/2^+[631]$,
$5/2^-[752]$ and the $5/2^+[633]$ neutron states are almost degenerate
across $N=138$ as shown in Ref.~\cite{Minkov24_PRC110}, therefore the
interpretation of a neutron configuration in a $N=138\pm 1$ nucleus as
a multi-particle--multi-hole excitation with respect to a
reference even-even nucleus with $N=138$ is ambiguous. Choosing $N =
140$ or $N = 142$ is thus better suited. Similarly the
$(1/2^+[640],1/2^-[530])$ proton doublet across $Z = 88$ in $^{226}$Ra
is found to be almost degenerate, but the energy difference between
these two $1/2$ states increases a bit when considering $^{228}$Ra and
heavier nuclei, so the above ambiguity disappears. We therefore retain
the following even-even nuclei:
\begin{itemize}
  \item \ce{^{228}_{88}Ra_{140}} to explore the orbitals of the
    $(3/2^+[631],3/2^-[741])$ of particle and hole characters with
    respect to $N=140$ together with the $1/2^-[530]$ proton particle
    state with respect to $Z = 88$;
  \item \ce{^{230}_{90}Th_{140}}, \ce{^{232}_{90}Th_{140}} and
    \ce{^{234}_{92}U_{142}} to study the $7/2^-[743]$ down-sloping
    neutron orbital as a particle state, the $3/2^+[631]$ up-sloping
    neutron orbital as a hole state and the above two $1/2$ proton
    orbitals as hole states in several odd-$A$ and odd-odd neighboring
    nuclei. 
\end{itemize}

\subsection{Octupole-deformation-energy curves of reference even-even
  nuclei}

\label{subsec_e-e_Q30_curves}

In the second step, we perform constrained Hartree--Fock--BCS
calculations for the selected even-even nuclei in their ground-state
configuration, with a constraint on the axial octupole moment of the
nucleon density as explained in
subsection~\ref{def_energy_curve}. This allows us to obtain the actual
behavior of single-particle energies with $\beta_3$, with which 
the estimates of the first step can be compared, and to select
relevant configurations to investigate the four scenarios of
subsection~\ref{subsec_scenarios}.

\begin{figure}[ht]
  \includegraphics[width=0.50\textwidth]{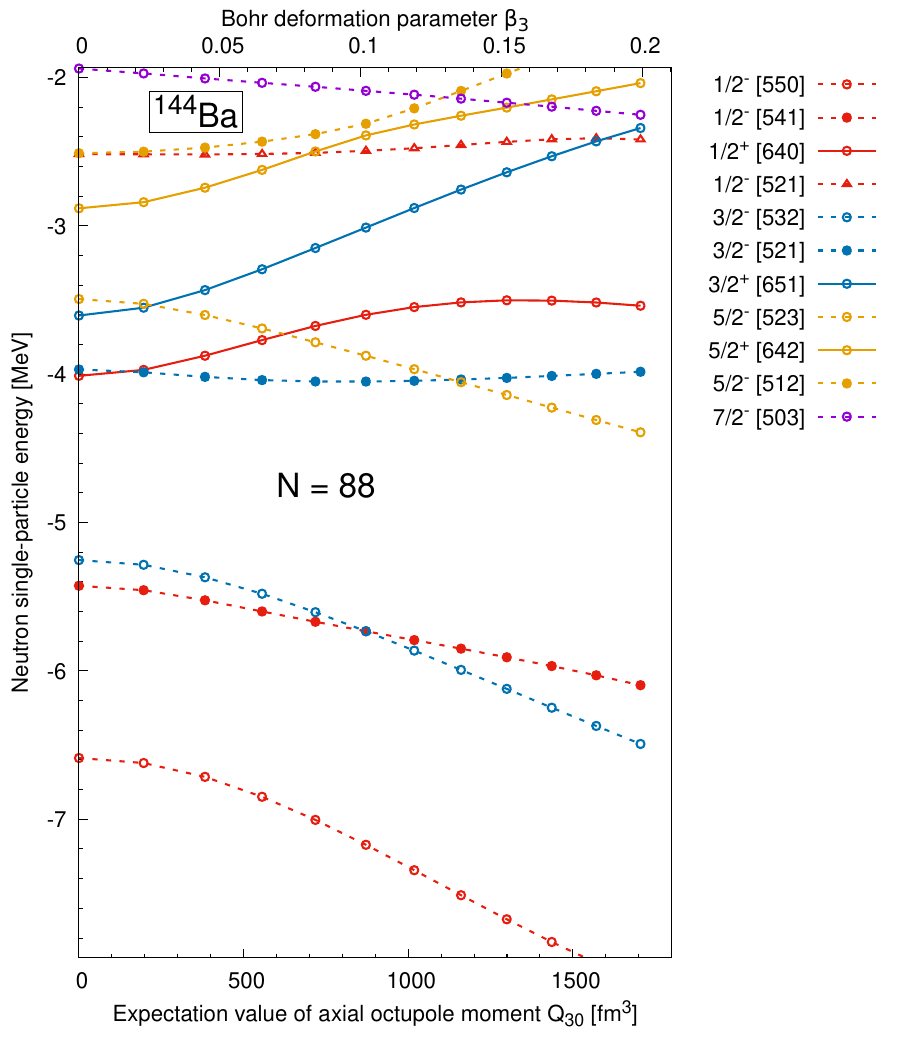} 
  \includegraphics[width=0.50\textwidth]{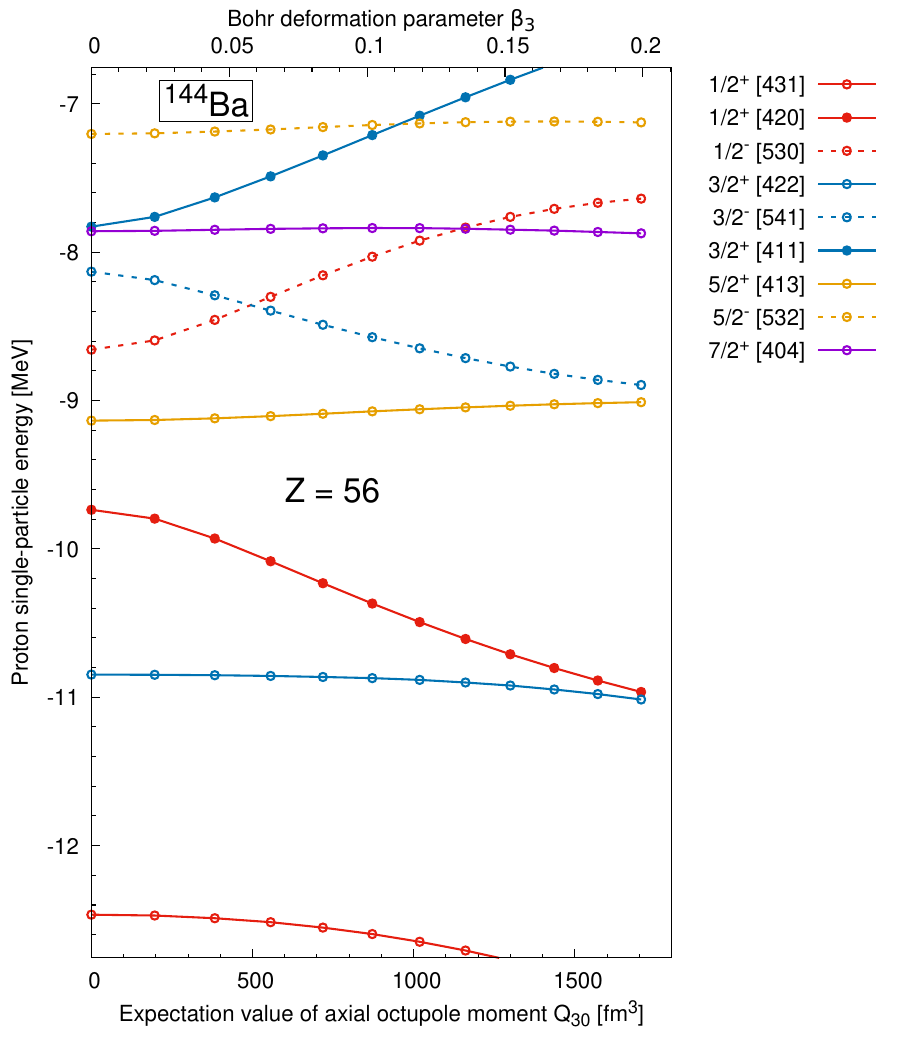}
  \caption{\label{fig_Ba144_Nilsson_diagrams} Neutron (left panel) and
    proton (right panel) single-particle energies as functions of
    axial octupole deformation of the \ce{^{144}_{56}Ba_{88}}
    nucleus. Labels correspond to the positive-$\Omega$ states of
    Kramers degenerate pairs.}
\end{figure}

\begin{figure}[ht]
  \includegraphics[width=0.50\textwidth]{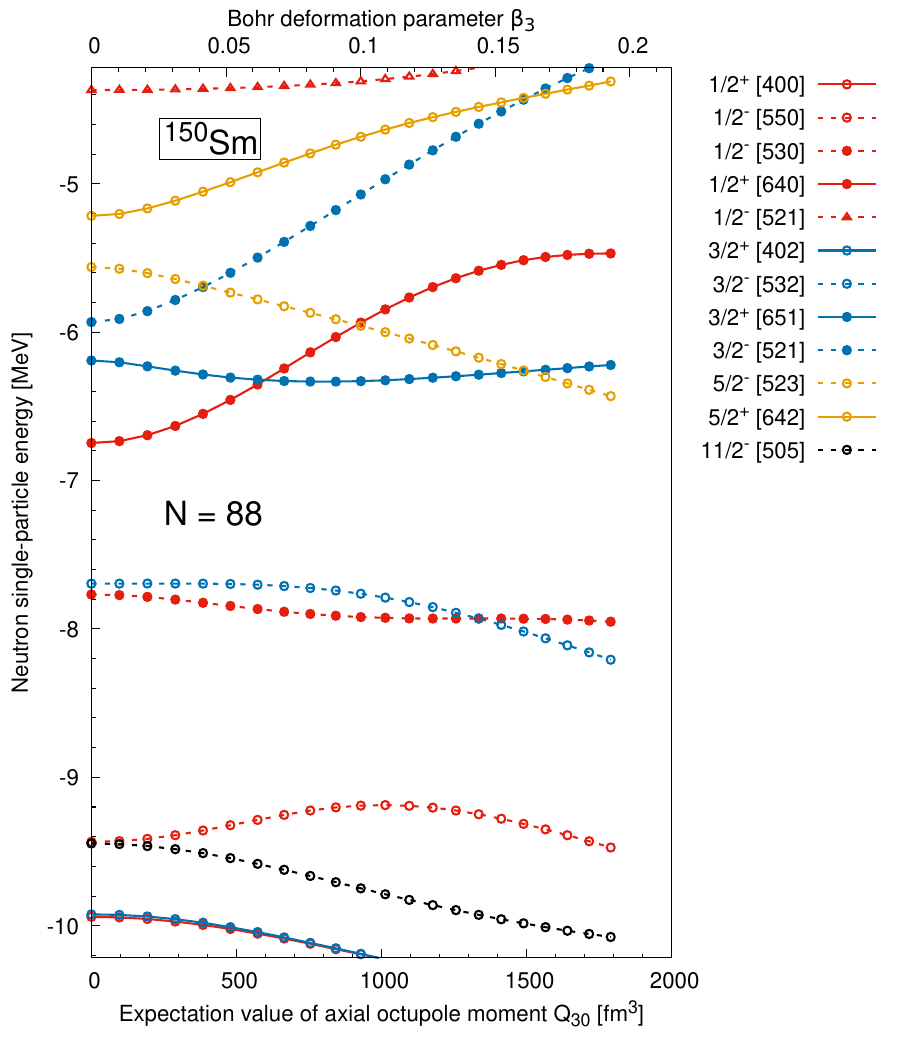} 
  \includegraphics[width=0.50\textwidth]{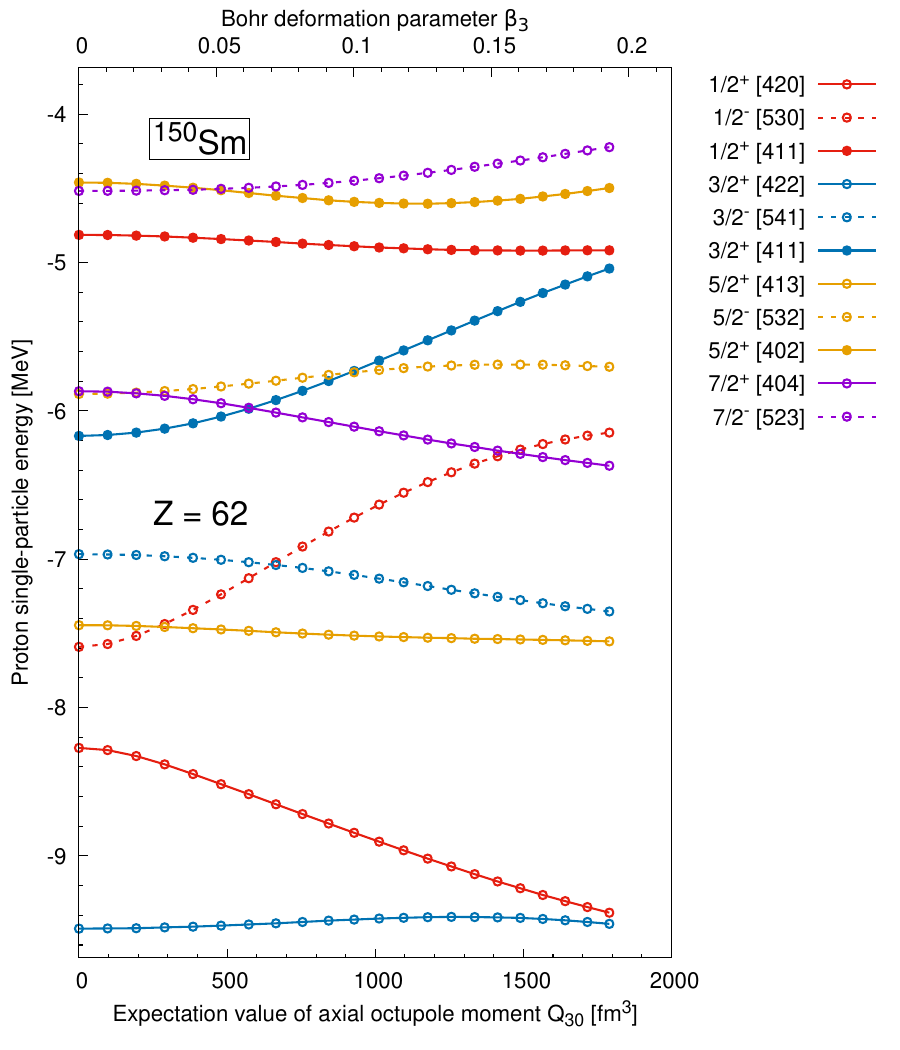}
  \caption{\label{fig_Sm150_Nilsson_diagrams} Same as
    figure~\ref{fig_Ba144_Nilsson_diagrams} for the
    \ce{^{150}_{62}Sm_{88}} nucleus.}
\end{figure}

\begin{figure}[ht]
  \includegraphics[width=0.50\textwidth]{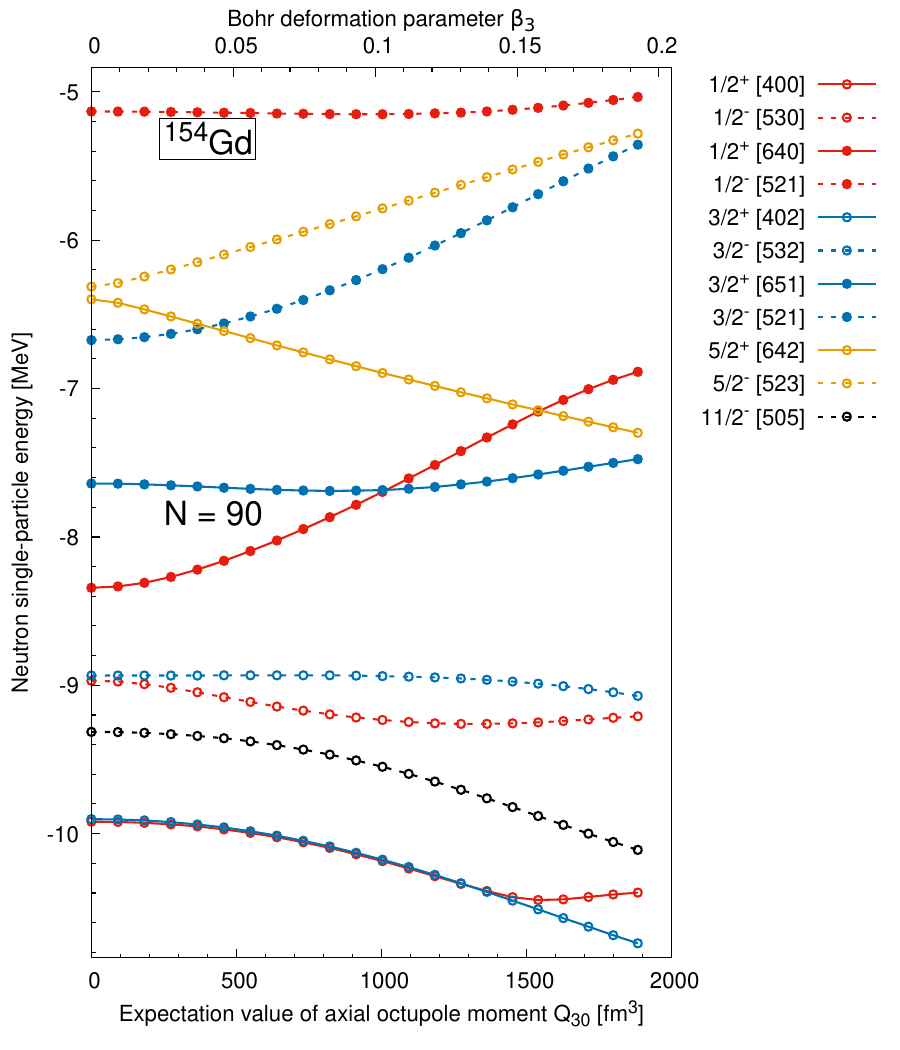} 
  \includegraphics[width=0.50\textwidth]{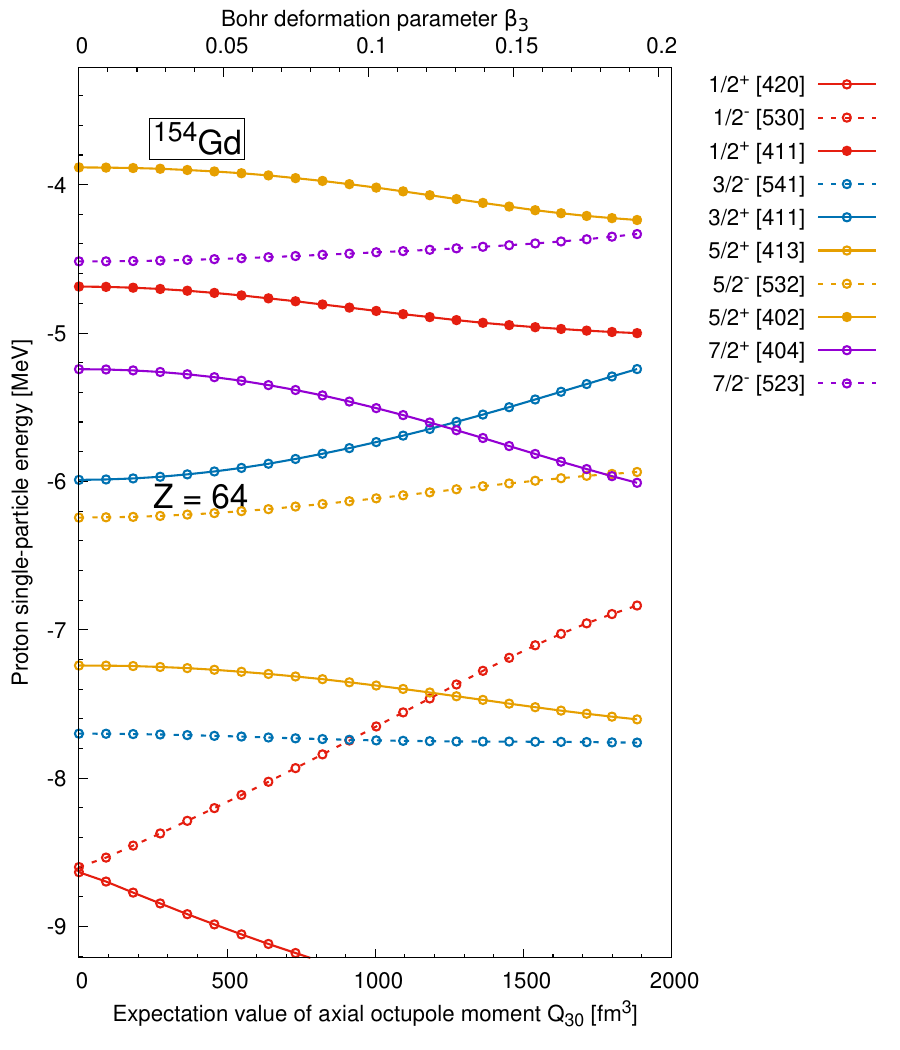}
  \caption{\label{fig_Gd154_Nilsson_diagrams} Same as
    figure~\ref{fig_Ba144_Nilsson_diagrams} for the
    \ce{^{154}_{64}Gd_{90}} nucleus.}
\end{figure}

\begin{figure}[ht]
  \includegraphics[width=0.50\textwidth]{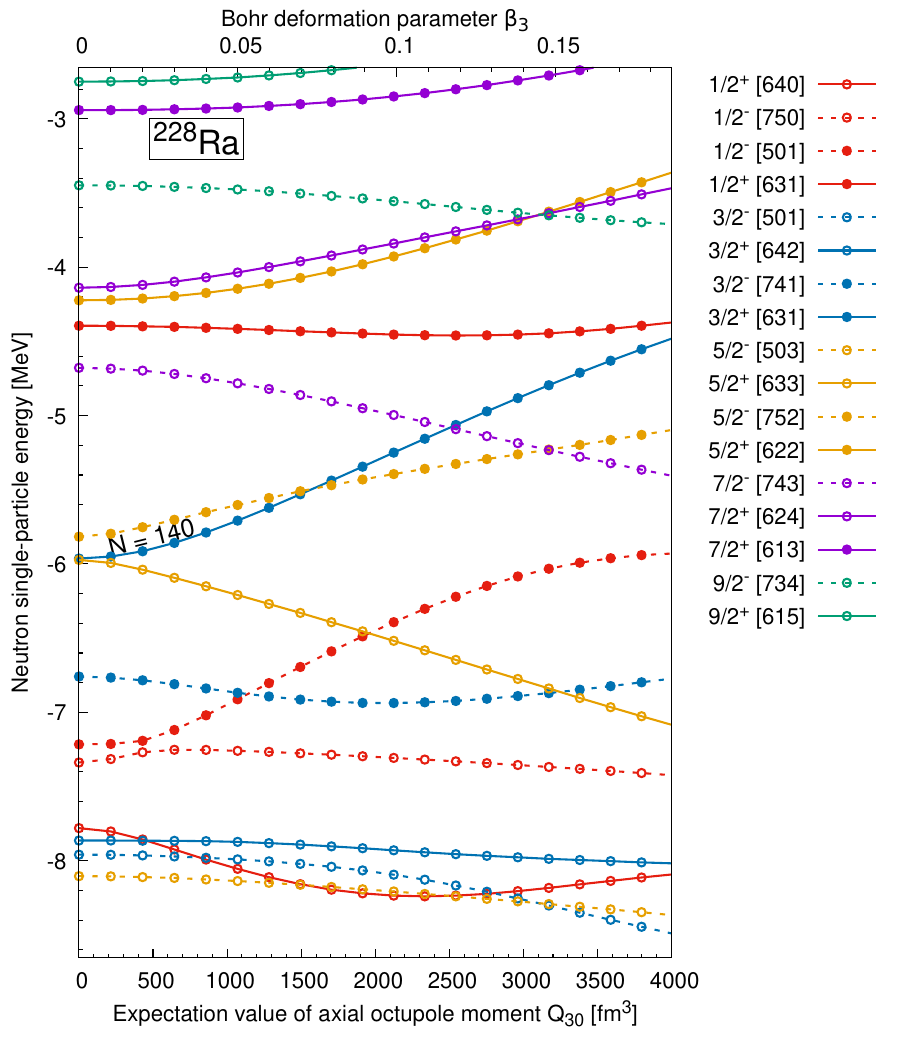} 
  \includegraphics[width=0.50\textwidth]{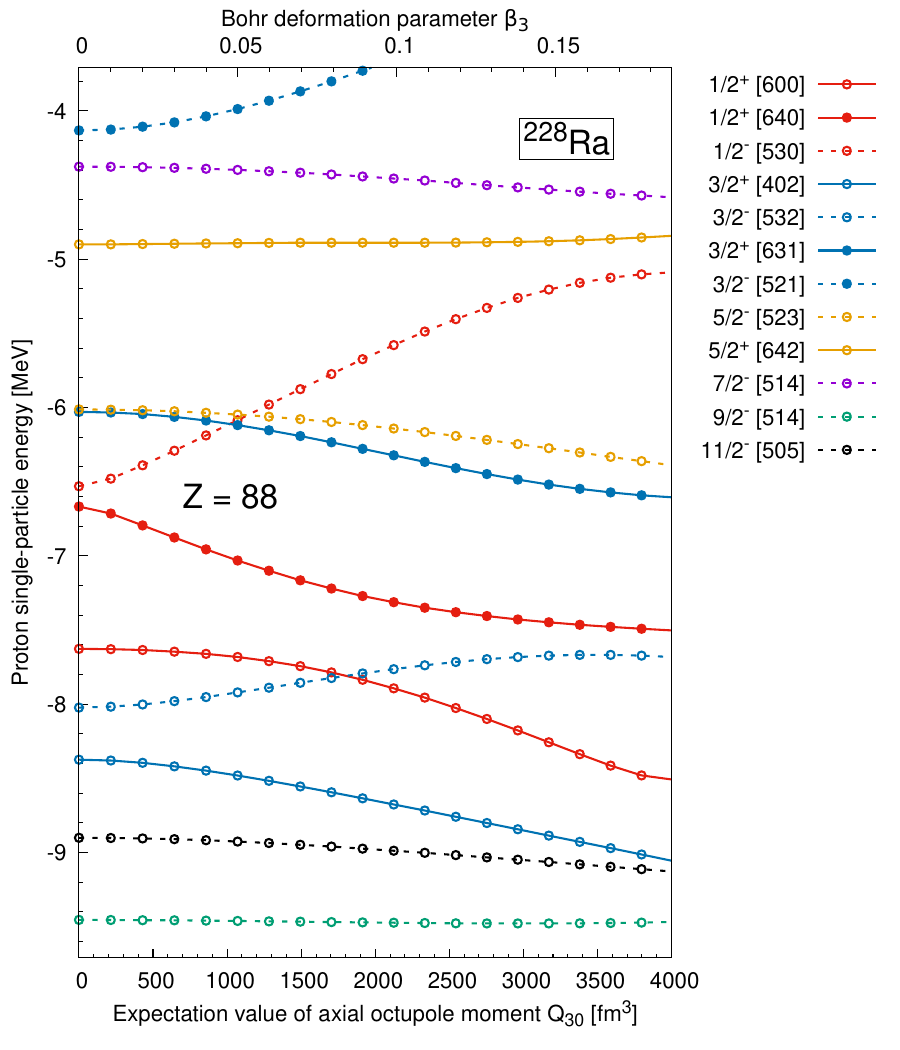}
  \caption{\label{fig_Ra228_Nilsson_diagrams} Same as
    figure~\ref{fig_Ba144_Nilsson_diagrams} for the
    \ce{^{228}_{88}Ra_{140}} nucleus.}
\end{figure}

\begin{figure}[ht]
  \includegraphics[width=0.50\textwidth]{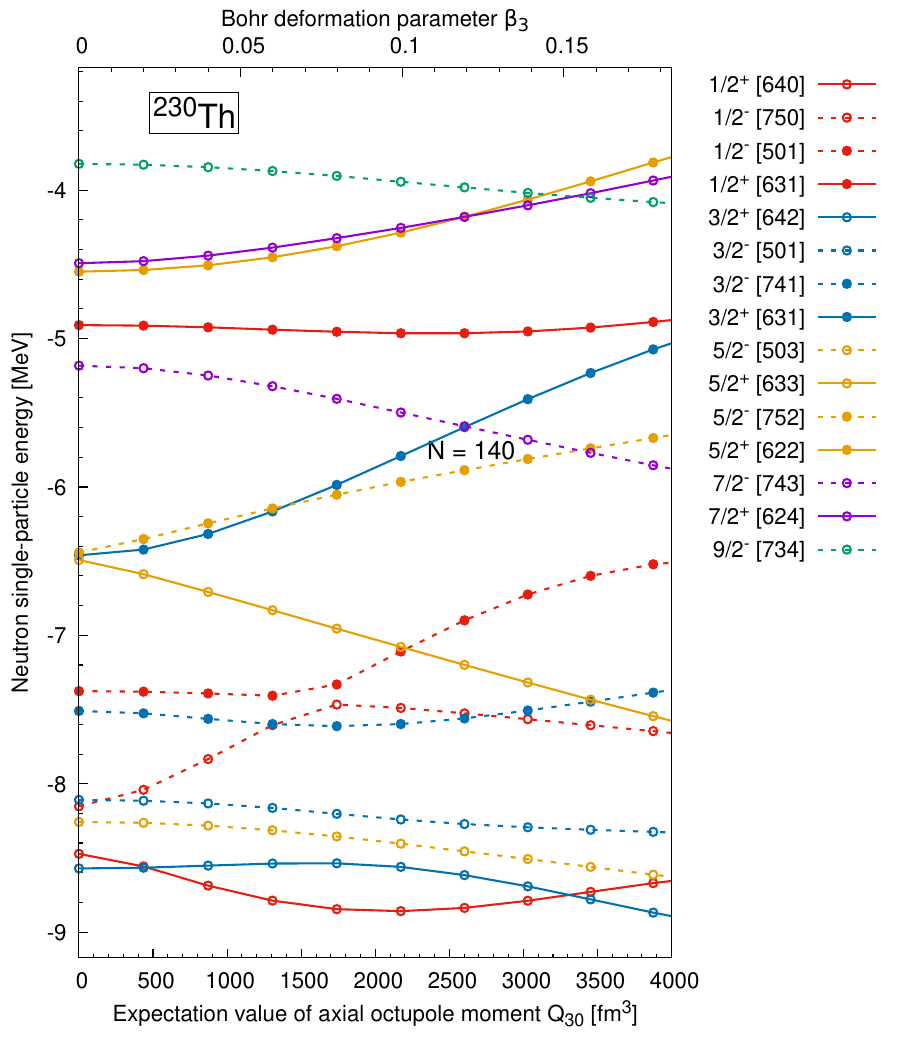} 
  \includegraphics[width=0.50\textwidth]{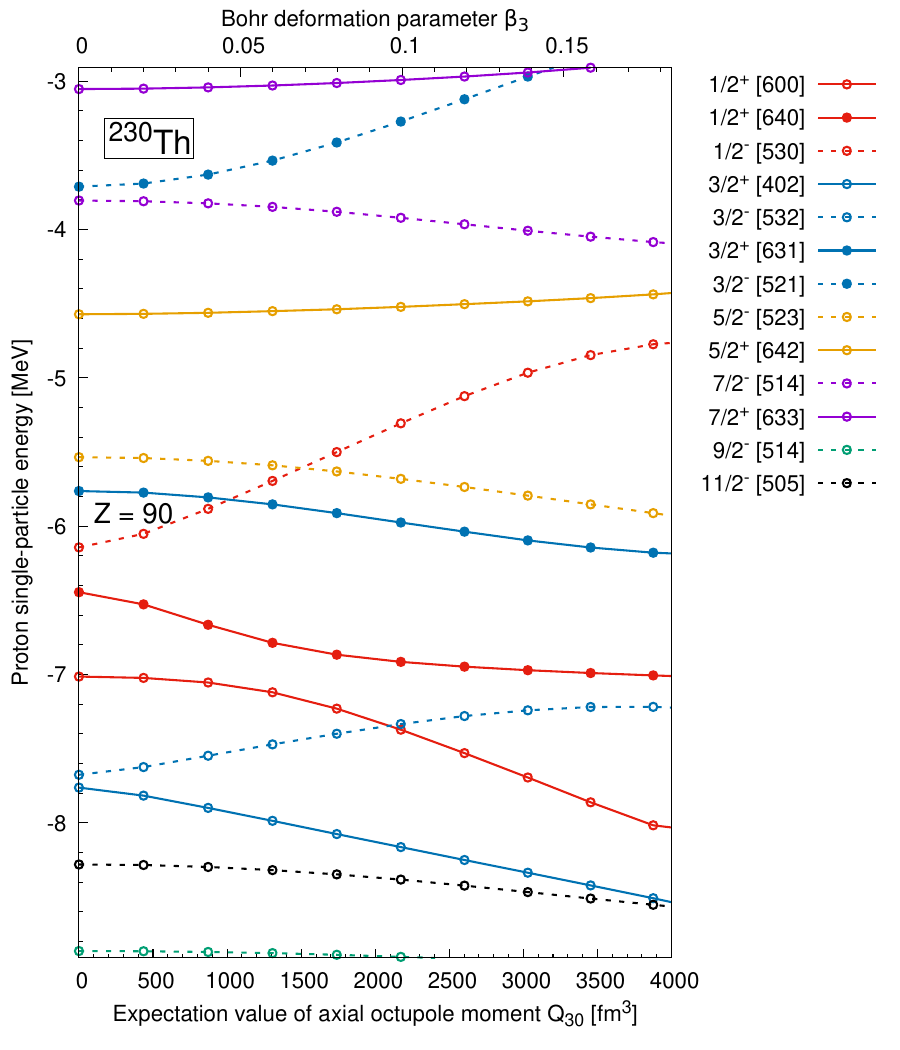}
  \caption{\label{fig_Th230_Nilsson_diagrams} Same as
    figure~\ref{fig_Ba144_Nilsson_diagrams} for the
    \ce{^{230}_{90}Th_{140}} nucleus.}
\end{figure}

\begin{figure}[ht]
  \includegraphics[width=0.50\textwidth]{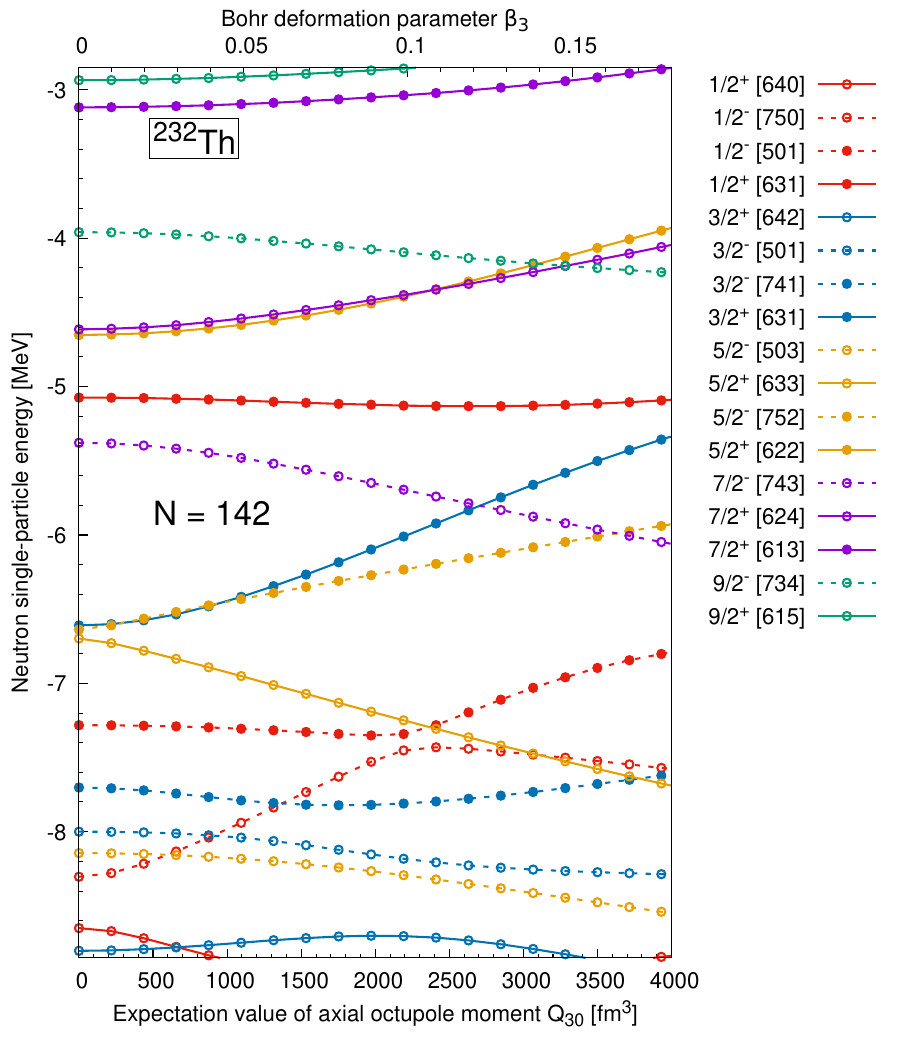} 
  \includegraphics[width=0.50\textwidth]{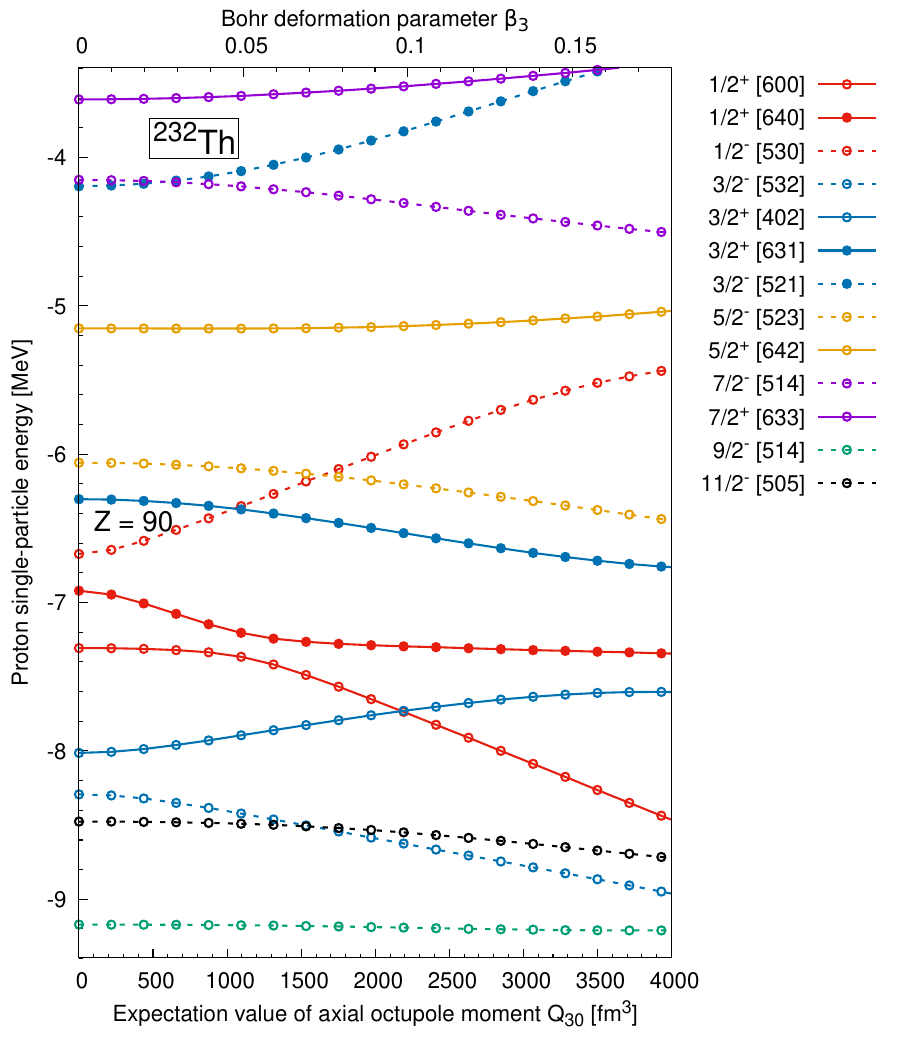}
  \caption{\label{fig_Th232_Nilsson_diagrams} Same as
    figure~\ref{fig_Ba144_Nilsson_diagrams} for the
    \ce{^{232}_{90}Th_{142}} nucleus.}
\end{figure}

\begin{figure}[ht]
  \includegraphics[width=0.50\textwidth]{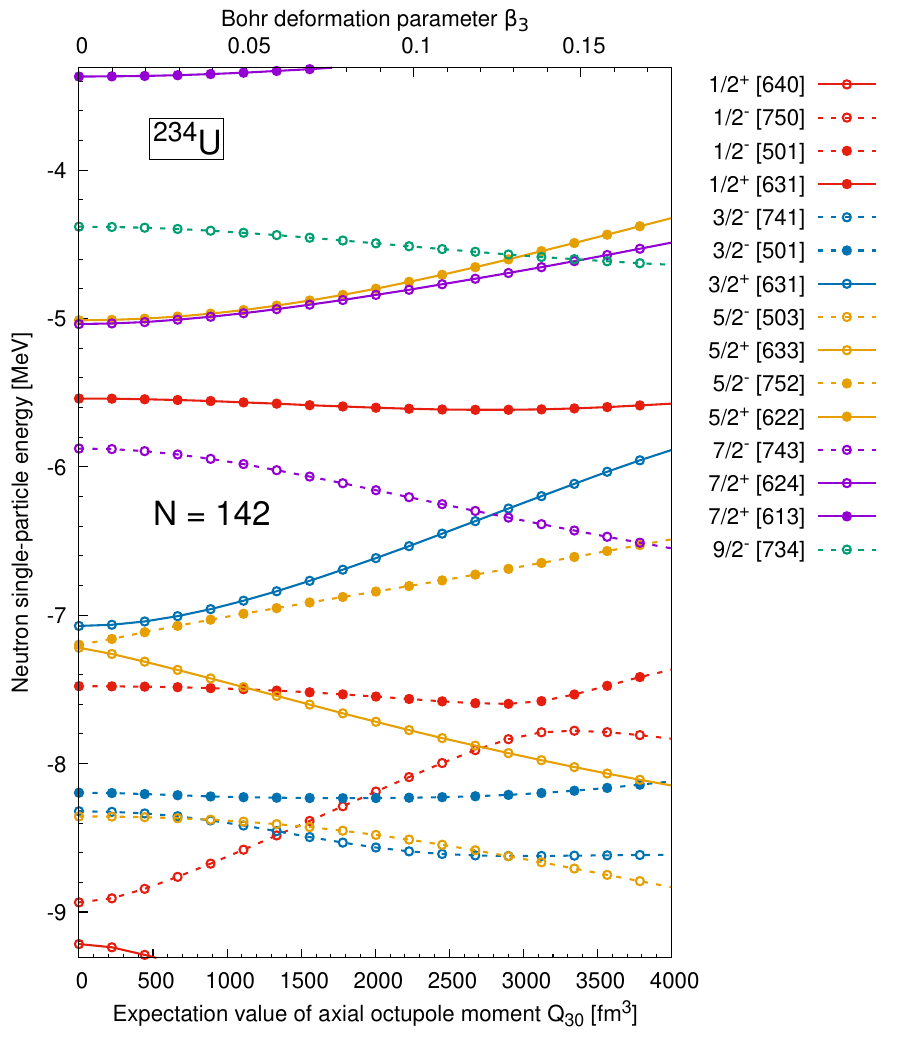} 
  \includegraphics[width=0.50\textwidth]{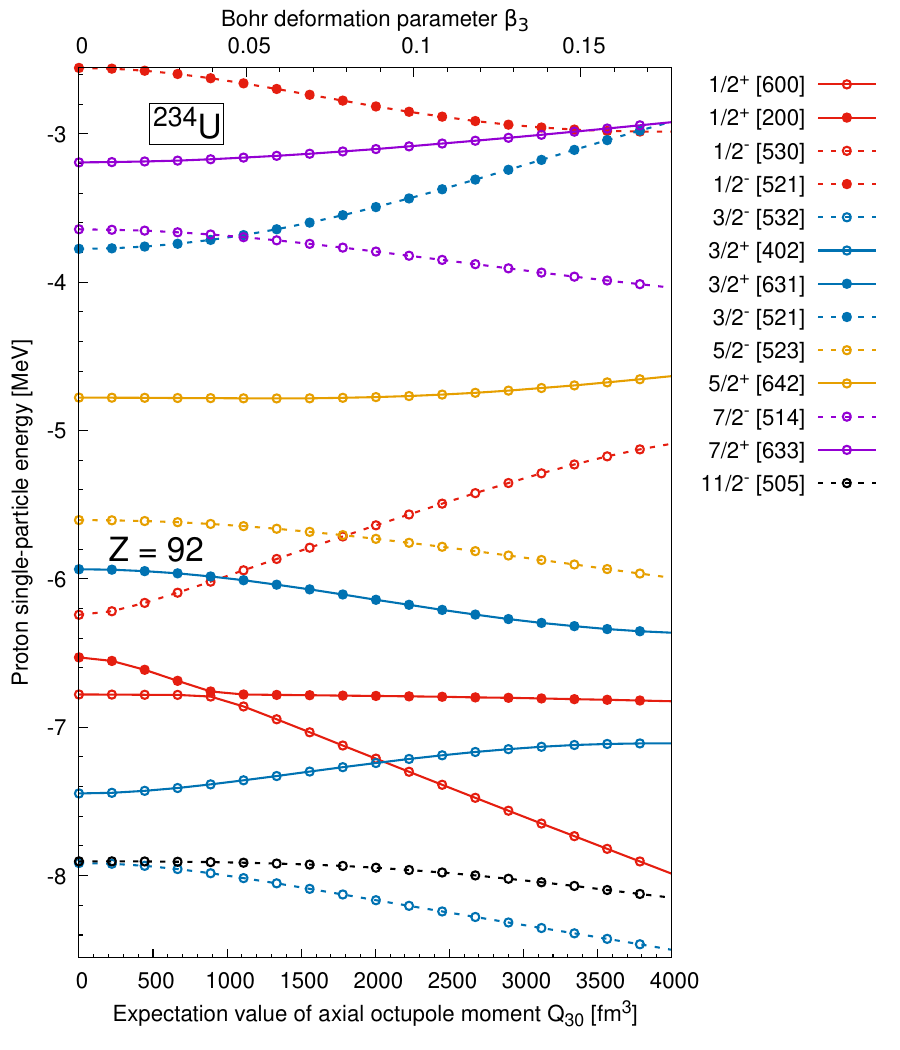}
  \caption{\label{fig_U234_Nilsson_diagrams} Same as
    figure~\ref{fig_Ba144_Nilsson_diagrams} for the
    \ce{^{232}_{90}Th_{142}} nucleus.}
\end{figure}

Figures \ref{fig_Ba144_Nilsson_diagrams} to
\ref{fig_U234_Nilsson_diagrams} show the resulting Nilsson diagrams
along the octupole-deformation-energy curves of the $^{144}$Ba,
$^{150}$Sm, $^{154}$Gd, $^{228}$Ra, $^{230}$Th, $^{232}$Th
and $^{234}$U nuclei, respectively. It is important to remember
that the Nilsson labels indicate the leading cylindrical
harmonic-oscillator contribution in a given single-particle state at
$\beta_3 = 0$. This contribution remains dominant also for small
values of $\beta_3$ but may become subleading at larger octupole
deformations. The curves of figures~\ref{fig_Ba144_Nilsson_diagrams} to
\ref{fig_U234_Nilsson_diagrams} are obtained by continuoulsy
following along the octupole-deformation axis the energies of the
single-particle states for each configuration.

As can be seen in figure~\ref{fig_Ba144_Nilsson_diagrams} for $^{144}$Ba and
figure~\ref{fig_Ra228_Nilsson_diagrams} for $^{228}$Ra,
the variations of single-particle
energies at small values of $\beta_3$ confirm the signs of most energy
corrections $\Delta e_x^{(\tau)}$ in
table~\ref{tab_Q30_sensitive_orbitals}. The discrepant cases
correspond to weakly octuple-sensitive orbitals.
Moreover, large absolute values of $\Delta e_x^{(\tau)} = e^{\prime (\tau)}_{x} -
e^{(\tau)}_{x}$ obtained for some doublets
$(\Omega^{\pi},\Omega^{-\pi})$ of single-particle states, for which the
leading Nilsson components obey the selection rules of $\widehat
Q_{30}$-coupling, correlate with strong repulsion of the corresponding
states in the Nilsson diagrams. That is especially the case of the
$(3/2^+[651], 3/2^-[521])$ neutron doublet near $N = 92$,
the $(3/2^+[411], 3/2^-[541])$ proton doublet near $Z = 62$ and
the $(1/2^+[640], 1/2^-[530])$ proton doublet near $Z = 88$.

In very few cases the perturbative model predicts an absolute
value of $\Delta e_x^{(\tau)}$ which overestimate the variation of the
corresponding single-particle energy as a function of $Q_{30}$ in the
Nilsson diagram. This is the case of the $3/2^-[521]$ neutron state
near $N=92$.

In contrast, the perturbative model predicts sizeable coupling between
the two $5/2$ neutron states of opposite parities near $N=138$ despite
different spin projections in their leading Nilsson
contributions. This is because of the subleading contributions which
are important in this case. However, this predicted coupling still
underestimates the strong repulsion observed in the Nilsson diagram of
neutron single-particle states in
figure~\ref{fig_Ra228_Nilsson_diagrams}. It turns out that these two
$5/2$ neutron states are almost degenerate in $^{228}$Th and
neighboring nuclei as already discussed in
Ref.~\cite{Minkov24_PRC110}. This causes problems of convergence at
small values of $\beta_3$ when blocking either of them as noticed in
paper I for the $K^{\pi} = 6^-$ $(7/2^-[743],5/2^+[633])_n$ isomeric
configuration of $^{234}$U. To solve these problems, a dedicated study
is called for.

\subsection{Selected configurations in odd-mass and odd-odd nuclei}

\label{subsec_selection}

From the above two steps, we have made a selection of relevant
configurations in odd-mass and odd-odd nuclei allowing to cover all
scenarios presented in subsection~\ref{subsec_scenarios} as well as
limiting cases between two scenarios. Table~\ref{tab_selection} shows
these selected nuclei and configurations, the reference
even-even nucleus, the lowest-energy multi-particle--multi-hole
excitation of this configuration with respect to the reference nucleus
in the Koopmans approximation and the expected scenario. Because of
the convergence problems mentioned in the previous paragraph, we have
discarded the two $5/2$ neutron states near $N=138$ in the blocked
configurations.

\begin{table}[t]
  \caption{Nuclei (ordered by increasing $Z$ and $A$)
    and configurations considered in this work, together
    with the reference even-even nucleus, the
    multi-particle--multi-hole excitation describing the configuration
    with respect to the reference nucleus, and the expected scenario.
    \label{tab_selection}}
  \centering
  \begin{tabular}{*{6}{c}}
    \hline
    \multirow{2}{*}{Nucleus} & \multirow{2}{*}{$K^{\pi}$} &
    Configuration & Reference 
    & mp-mh & Expected  \\
    & & at $\beta_3 = 0$ & nucleus & excitation & scenario \\
    \hline
    \ce{^{145}_{56}Ba_{89}}
    & $3/2^-$ & $(3/2^- [521])_n$ &
    \multirow{2}{*}{\ce{^{144}_{56}Ba_{88}}} & $\rm(1p0h)_n$ & (b) \\
    & $5/2^-$ & $(5/2^- [523])_n$ & & $\rm(1p0h)_n$ & (b) \\
    \hline
    \ce{^{151}_{63}Eu_{88}}& $5/2^-$ & $(5/2^- [532])_p$ &
    \multirow{3}{*}{\ce{^{150}_{62}Sm_{88}}} & $\rm(1p0h)_n$ & (b) \\
    & $3/2^+$ & $(3/2^+ [411])_p$ & & $\rm(1p0h)_n$ & (b) or (d) \\
    & $7/2^+$ & $(7/2^+ [404])_p$ & & $\rm(1p0h)_n$ & (b) \\
    \hline
    \ce{^{152}_{63}Eu_{89}}
    & $3^-$ & $(1/2^+ [640])_n (5/2^- [532])_p$ &
    \multirow{3}{*}{\ce{^{150}_{62}Sm_{88}}} & $\rm(1p0h)_n(1p0h)_p$ & (b) or (d) \\
    & $2^+$ & $(1/2^+ [640])_n (3/2^+ [411])_p$ & & $\rm(1p0h)_n(1p0h)_p$ & (d) \\
    & $5^+$ & $(3/2^+ [651])_n (7/2^+ [404])_p$ & & $\rm(1p0h)_n(1p0h)_p$ & (b) \\
    \hline
    \ce{^{154}_{63}Eu_{91}}
    & $5^+$ & $(3/2^+ [651])_n (7/2^+ [404])_p$ &
    \multirow{2}{*}{\ce{^{154}_{64}Gd_{90}}} & $\rm(1p0h)_n(1p2h)_p$ & (c) \\
    & $4^+$ & $(1/2^+ [640])_n (7/2^+ [404])_p$ & & $\rm(2p1h)_n(1p2h)_p$ & (a) \\
    \hline
    \ce{^{226}_{87}Fr_{139}}
    & $2^+$ & $(3/2^- [741])_n (1/2^- [530])_p$ &
    \multirow{2}{*}{\ce{^{228}_{88}Ra_{140}}} & $\rm(0p1h)_n(1p2h)_p$
    & (a) or (d) \\
    & $2^-$ & $(3/2^+ [631])_n (1/2^- [530])_p$ & &
    $\rm(0p1h)_n(1p2h)_p$ & (a) or (d) \\
    \hline
    \ce{^{228}_{89}Ac_{139}}
    & $2^-$ & $(3/2^+ [631])_n (1/2^- [530])_p$ &
    \multirow{2}{*}{\ce{^{228}_{88}Ra_{140}}} & $\rm(0p1h)_n(1p0h)_p$ & (b) or (c) \\
    & $3^+$ & $(3/2^+ [631])_n (3/2^+ [631])_p$ & &
    $\rm(0p1h)_n(1p0h)_p$ & (a) or (d) \\
    \hline
    \ce{^{231}_{89}Ac_{142}}
    & $1/2^+$ & $(1/2^+ [640])_p$ &
    \multirow{3}{*}{\ce{^{232}_{90}Th_{142}}} & $\rm(0p1h)_p$ & (a) \\
    & $11/2^-$ & $(5^-)_n (1/2^+ [640])_p$ & & $\rm(1p1h)_n(0p1h)_p$ & (a)
    or (c) \\
    & $11/2^+$ & $(5^-)_n (1/2^- [530])_p$
    & & $\rm(1p1h)_n(0p1h)_p$ & (c) \\
    \multicolumn{3}{r}{with $(5^-)_n=(7/2^-[743],3/2^+[631])_n$} & \\
    \hline
    \ce{^{229}_{90}Th_{139}}
    & $3/2^+$ & $(3/2^+ [631])_n$ &
    \multirow{2}{*}{\ce{^{230}_{90}Th_{140}}} & $\rm(0p1h)_n$ & (b) \\
    & $7/2^-$ & $(7/2^- [743])_n$ & & $\rm(1p2h)_n$ & (b) \\
    \hline
    \ce{^{230}_{91}Pa_{139}}
    & $2^-$ & $(3/2^+ [631])_n (1/2^- [530])_p$ &
    \ce{^{230}_{90}Th_{140}} & $\rm(0p1h)_n(2p1h)_n$ & (b) or (c) \\
    \hline
    \ce{^{234}_{91}Pa_{143}}
    & $4^+$ & $(7/2^- [743])_n (1/2^- [530])_p$ &
    \ce{^{234}_{92}U_{142}} & $\rm(1p0h)_n(0p1h)_p$ & (c) \\
    \hline
    \ce{^{235}_{92}U_{143}} & $7/2^-$ & $(7/2^- [743])_n$ &
    \multirow{2}{*}{\ce{^{234}_{92}U_{142}}} & $\rm(1p0h)_n$ & (a) or (c) \\
    & $3/2^+$ & $(3/2^+ [631])_n$ & & $\rm(2p1h)_n$ & (c) \\
    \hline
  \end{tabular}
\end{table}

We present the octupole-deformation energy curves and compare the
expected scenarios with the actual ones for the $^{145}$Ba and
$^{151,152,154}$Eu nuclei in the next subsection, 
whereas those of the $^{226}$Fr, $^{228,231}$Ac, $^{230,234}$Pa and $^{235}$U nuclei are
presented and discussed in subsection~\ref{subsec_actinides}.

%
%
%
%

\section{Results and discussion}

\label{sec_res}

\subsection{Octupole deformation in some odd-mass and odd-odd nuclei near $A=150$}

\label{subsec_rare_earths}

First we focus on odd-mass nuclei (1-quasiparticle configurations)
then we will deal with doubly-odd nuclei (2-quasiparticle configurations).

\paragraph{Odd-mass $^{145}$Ba and $^{151}$Eu nuclei.}

As can be seen on the left panel of figure~\ref{fig_cutQ30_Ba144_Ba145},
\begin{figure}[ht]
  \centering
  \includegraphics[width=\textwidth]{figures/fig_cutQ30_Ba144_Ba145.pdf}
\caption{Deformation energy curves along the $Q_{30}$ direction in the
  $K^{\pi} = 3/2^- ([521])_n$ and $K^{\pi} = 5/2^- ([523])_n$
  configurations of $^{145}$Ba (right panel) and in the ground-state
  configuration of the even-even reference nucleus $^{144}$Ba (left
  panel). \label{fig_cutQ30_Ba144_Ba145}}
\end{figure}
\begin{figure}[ht]
  \centering
  \includegraphics[width=\textwidth]{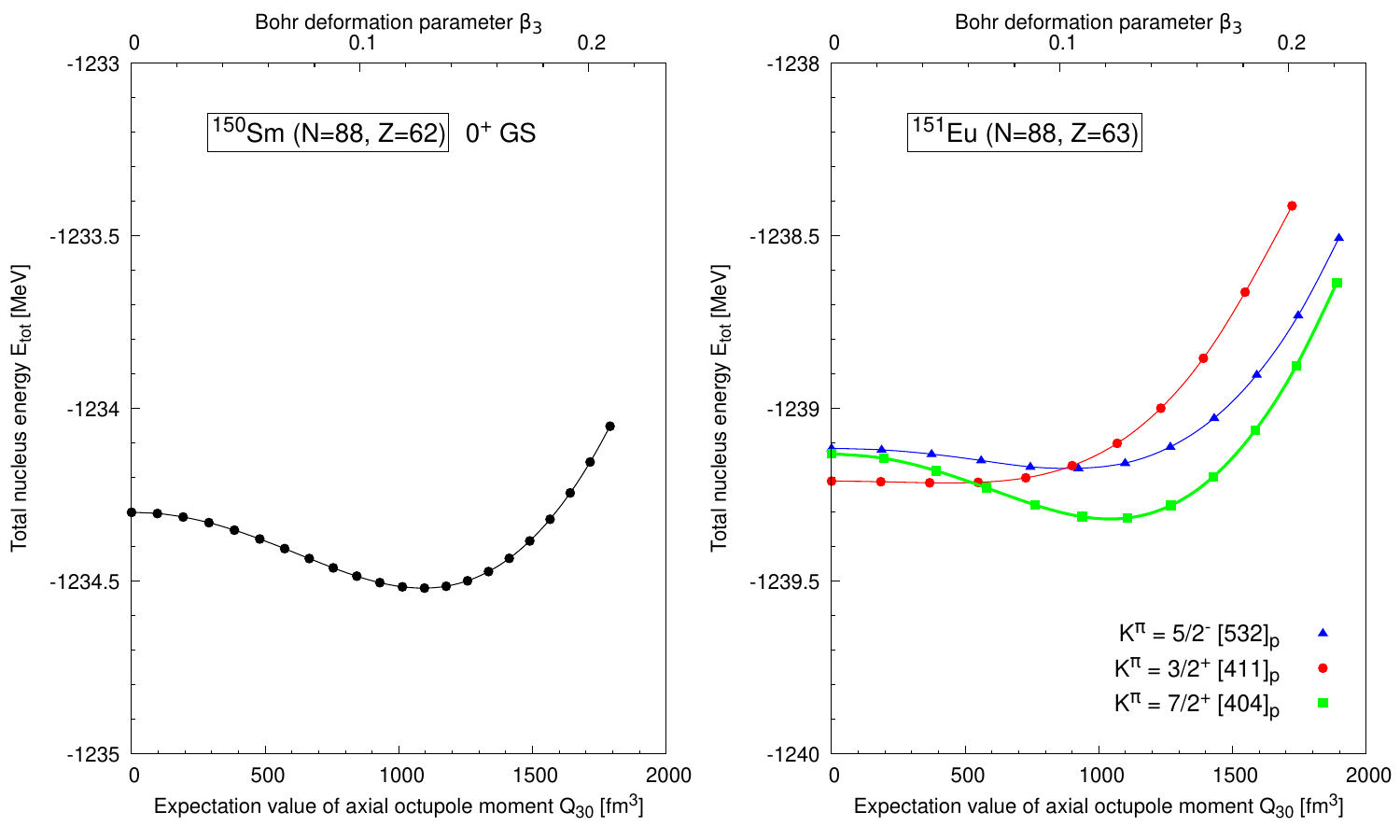}
\caption{Same as figure~\ref{fig_cutQ30_Ba144_Ba145} for the $^{151}$Eu
  nucleus and the reference nucleus $^{150}$Sm. \label{fig_cutQ30_Sm150_Eu151}}
\end{figure}
the constrained Hartree--Fock--BCS calculations in the ground-state
configuration of the even-even $^{144}$Ba nucleus produce an
octupole-deformation-energy curve  with a 
local minimum at a finite value of $Q_{30}$. This means that octupole
deformation is energetically favorable in the ground-state
configuration of this even-even nucleus ($\Delta E_0 < 0$). Moreover, the
$\mathcal C_1 = 5/2^- [523]$ neutron configuration of $^{145}$Ba
corresponds to the creation of one neutron state in the $^{144}$Ba
reference nucleus. Therefore, according to Eqs.~(\ref{E_Phi'_perturb}) to
(\ref{Delta_en}), the perturbed energy $E_{\mathcal C}[\Phi']$ in the
configuration $\mathcal C_1$ of the $^{145}$Ba nucleus is approximately
\eq{
E_{\mathcal C_1}[\Phi'] = E_{\mathcal C_1}[\Phi] + \Delta E_0 + \Delta
e^{(\rm n)}_{5/2^- [523]}
}
with $\Delta E_0 < 0$ and $\Delta e^{(\rm n)}_{5/2^- [523]} < 0$.
One thus expects that $E_{5/2^-[523]}[\Phi']
< E_{5/2^-[523]}[\Phi]$, which means that at small $\beta_3$ values,
an axial octupole shape is energetically favorable in the
$5/2^-$~$([523])_p$ configuration of the $^{145}$Ba nucleus
(scenario~(b)). This agrees with the actual octupole-constrained
Hartree--Fock--BCS calculations.

Similarly to the configuration $\mathcal C_1$, the
neutron configuration $\mathcal C_2 = 3/2^-[521]$ of $^{145}$Ba is of
particle character, hence 
\eq{
E_{\mathcal C_2}[\Phi'] = E_{\mathcal C_2}[\Phi] + \Delta E_0 + \Delta
e^{(\rm n)}_{3/2^- [521]}
}
with $\Delta e_{3/2^-[521]}^{(\rm n)} < 0$. Therefore
$\Delta E_0 + \Delta e_{3/2^-[521]}^{(\rm n)} < \Delta E_0$ is expected
to produce a deeper local minimum at finite $\beta_3$ in the
deformation energy curve of $^{145}$Ba than the local minimum in the
deformation energy curve of $^{144}$Ba. This is indeed the case in
actual Hartree--Fock-BCS calculations (scenario~(b)) as can be deduced
from the same scales used for both nuclei in
figure~\ref{fig_cutQ30_Ba144_Ba145}.

For the $^{151}$Eu nucleus, figure~\ref{fig_cutQ30_Sm150_Eu151} shows
that the selected configurations span a range of
scenarios of type~(b) (for the $K^{\pi} = 7/2^+$ and $K^{\pi} = 5/2^-$
configurations), up to the limiting case between scenarios~(c)
and (d) for the $K^{\pi} = 3/2^+$ configuration. The corresponding
blocked proton states are all particle states with respect to the
ground-state configuration of the $^{150}$Sm reference nucleus, as can
been from its Nilsson diagrams in figure~\ref{fig_Sm150_Nilsson_diagrams}.

The $K^{\pi} = 3/2^+$ limiting case can be understood from the rather
large and positive value of $\Delta e^{(\rm p)}_{3/2^+[411]}$
calculated in $^{150}$Sm, namely about 90~keV, as compared to the
value of about $-20$~keV for $\Delta E_0$ when calculated as the
difference in total energy at $\beta_3 = 0$ and $\beta_3 = 0.02$. It
is worth noting that the value of $\Delta e^{(\rm p)}_{3/2^+[411]}$
calculated in $^{144}$Ba and reported in
table~\ref{tab_Q30_sensitive_orbitals} is even much larger than the
above 90~keV, whereas the actual variation of the energy of the
$3/2^+[411]$ proton state between $\beta_3 = 0$ and $\beta_3 = 0.02$
is calculated to be about 20~keV in $^{150}$Sm (hence $\Delta
E_0+\Delta e \approx 0$ as suggested by
figure~\ref{fig_cutQ30_Sm150_Eu151}). The use of the 
estimated $\Delta e_x^{(\tau)}$ values should thus be made only as a
guide to interprete the trends in actual deformation energy curves. In
the subsequent discussions, we shall thus use the actual $\Delta
e_x^{(\tau)}$ values deduced from constrained Hartree--Fock--BCS
calculations.

In contrast, the values of $\Delta e^{(\rm
  p)}_{5/2^-[532]}$ and $\Delta e^{(\rm p)}_{7/2^+[404]}$ from the
deformation energy curve of $^{150}$Sm are about 10~keV and $-15$~keV,
respectively. This could explain why the local minimum of the $K^{\pi} = 5/2^-$
deformation energy curve is less deep than the one of the $K^{\pi} = 7/2^+$
deformation energy curve. However the negative value of
$\Delta e^{(\rm p)}_{7/2^+[404]}$ would suggest a local minimum deeper
than the one in the $^{150}$Sm deformation energy curve, at variance
with figure~\ref{fig_cutQ30_Sm150_Eu151} probably because of
nonperturbative effects.

\paragraph{Odd-odd $^{152,154}$Eu nuclei.}

\begin{figure}[ht]
  \centering
  \includegraphics[width=\textwidth]{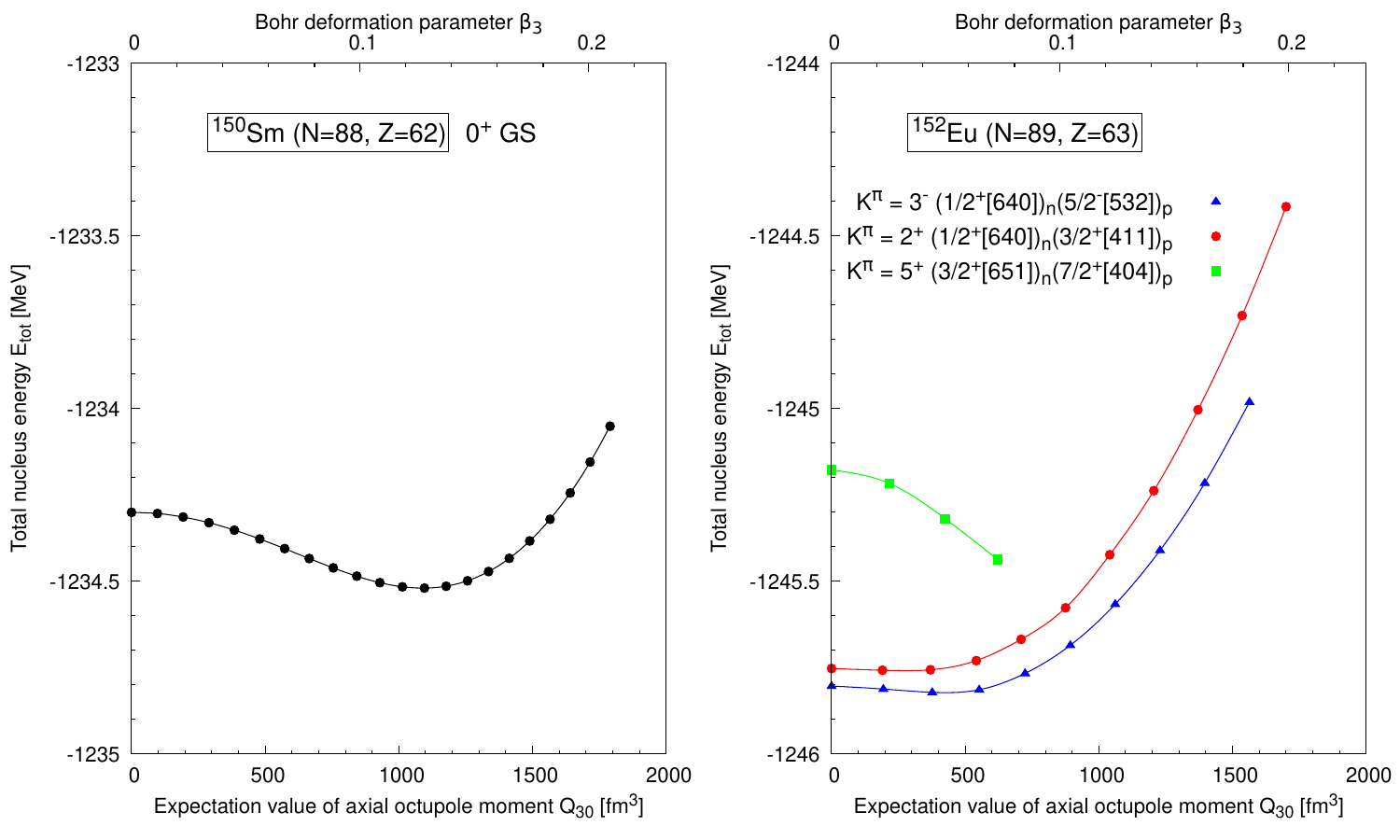}
\caption{Same as figure~\ref{fig_cutQ30_Ba144_Ba145} for the $^{152}$Eu
  nucleus and the reference nucleus $^{150}$Sm. \label{fig_cutQ30_Sm150_Eu152}}
\end{figure}

The octupole-deformation-energy curves of the three studied
configurations of $^{152}$Eu are plotted in
figure~\ref{fig_cutQ30_Sm150_Eu152}. 
As already noted, the
ground-state minimum of the reference even-even nucleus $^{150}$Sm
has a finite octupole deformation.

In the $K^{\pi} = 3^-$
$(1/2^+[640])_n (5/2^-[532])_p$ and $K^{\pi} = 2^+$ $(1/2^+[640])_n
(3/2^+[411])_p$ configurations, the $^{152}$Eu nucleus is very soft 
against octupole deformations, which corresponds to the limiting case
between scenarios~(b) and (d). This can be explained by the fact that
these two configurations are obtained by blocking the $1/2^+[640]$ neutron
state above $N=88$ and that the energy of this orbital increases with
$\beta_3$, so that $\Delta e^{(\rm n)}_{1/2^+[640]} > 0$. At the same
time, on the proton side, the blocked proton state for both
configurations is also above the Fermi level of $^{150}$Sm and
increases with $\beta_3$. One thus obtains $\Delta e^{(\rm
  n)}_{1/2^+[640]} + \Delta e^{(\rm p)}_{5/2^-[532]} > 0$ and $\Delta e^{(\rm
  n)}_{1/2^+[640]} + \Delta e^{(\rm p)}_{3/2^+[411]} > 0$, therefore
$\Delta e$ at least partially compensates $\Delta E_0 < 0$ in
$^{150}$Sm: this gives scenario (b) if this compensation is not enough
to overcome $\Delta E_0$ or scenario (d) if $\Delta e > \big|\Delta
E_0\big|$. 

In contrast in the $K^{\pi} = 5^+$ $(3/2^+[651])_n (7/2^+[404])_p$
configuration of $^{152}$Eu the blocked neutron and proton states, of particle
character with respect to the ground-state configuration of
$^{150}$Sm, have energies decreasing with $\beta_3$, hence $\Delta e <
0$. Given that $\Delta E_0 < 0$, we thus have the net effect $\Delta E_0 + \Delta
e < \Delta E_0$. This explains why the deformation energy curve of
this configuration in figure~\ref{fig_cutQ30_Sm150_Eu152} is
down-sloping for small values of $\beta_3$, which suggests a local
minimum at finite octupole deformation although calculations did not
converge beyond $Q_{30} \sim 600$~$\rm fm^3$. This situation falls
into scenario~(b).

\begin{figure}[ht]
  \includegraphics[width=\textwidth]{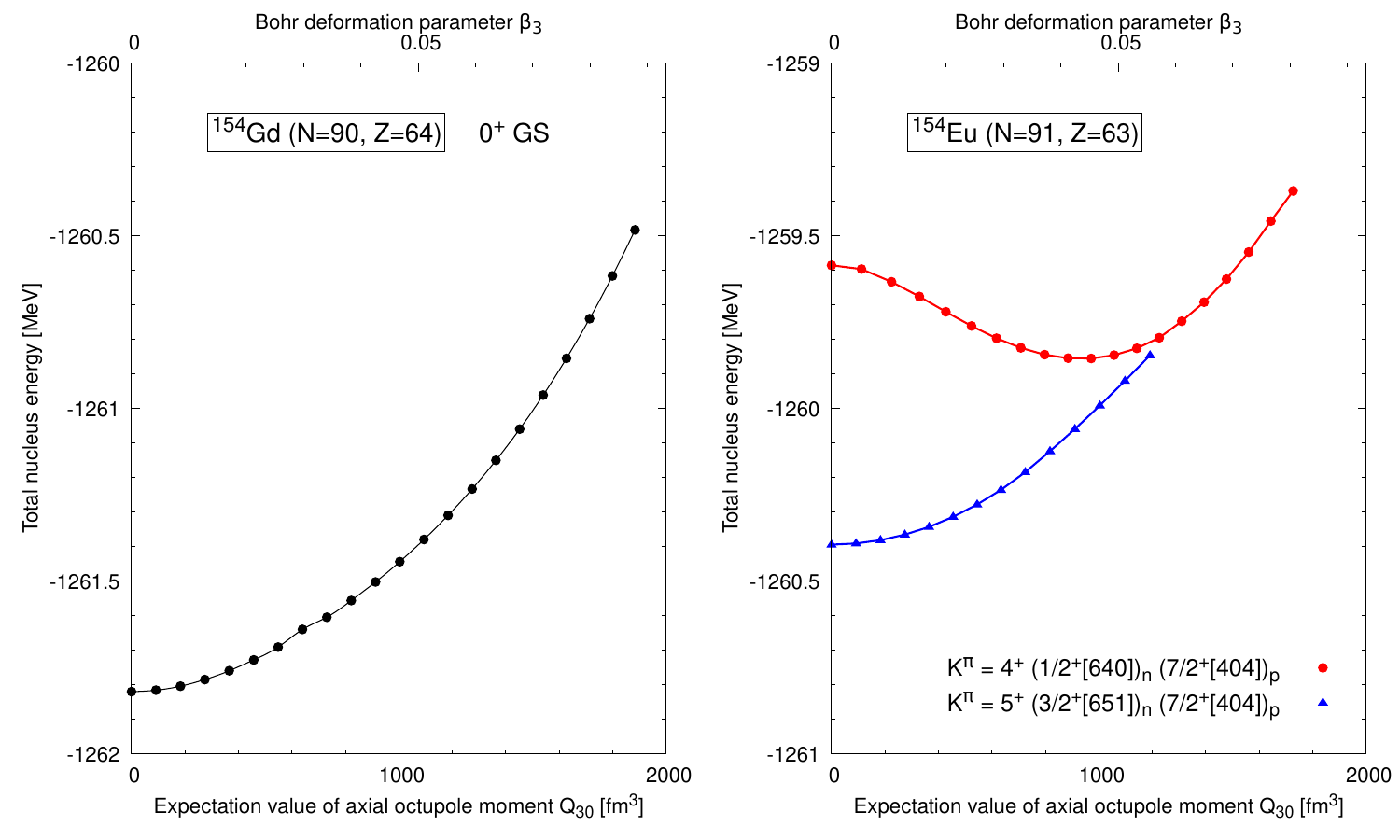}
\caption{Same as figure~\ref{fig_cutQ30_Ba144_Ba145} for the $^{154}$Eu
  nucleus and the reference nucleus $^{154}$Gd. \label{fig_cutQ30_Gd154_Eu154}}
\end{figure}

Regarding the $^{154}$Eu nucleus, its reference even-even nucleus
$^{154}$Gd has a left-right symmetric shape in its ground
state as can be seen from figure~\ref{fig_cutQ30_Gd154_Eu154}.
The Nilsson diagrams of $^{154}$Gd are plotted in
figure~\ref{fig_Gd154_Nilsson_diagrams}.
The $K^{\pi} = 4^+$ $(1/2^+[640])_n (7/2^+[404])_p$ configuration of
$^{154}$Eu is described as a neutron 2-particle--1-hole, proton
1-particle--2-hole excitation with respect to the ground-state
configuration of $^{154}$Gd. The lowest-energy of a neutron excitation
of this order involving $1/2^+[640]$ as the blocked state is obtained
by creating a pair of neutrons on the $3/2^+[651]$ level and
annihilating the conjugate state of the $1/2^+[640]$ neutron
hole. Similarly on the proton side, the lowest-energy realization of a
1-particle--2-hole excitation with $7/2^+[404]$ as the blocked state
is the annihilation of the protons on the $5/2^-[523]$ level and the
creation of a proton on the $7/2^+[404]$ level. Therefore the total
single-particle energy correction $\Delta e(4^+)$ is given by
\eq{
  \Delta e(4^+) = 2\,\Delta e^{(\rm n)}_{3/2^+[651]} - \Delta e^{(\rm
    n)}_{1/2^+[640]} + \Delta e_{7/2^+[404]}^{(\rm p)} - 2\,\Delta
  e^{(\rm p)}_{5/2^-[523]} \,.
}
According to figure~\ref{fig_Gd154_Nilsson_diagrams}, we have $\Delta
e^{(\rm n)}_{3/2^+[651]} \lesssim 0$, $\Delta e^{(\rm n)}_{1/2^+[640]}
> 0$, $\Delta e_{7/2^+[404]}^{(\rm p)} < 0$ and $\Delta
  e^{(\rm p)}_{5/2^-[523]} \gtrsim 0$, therefore $\Delta e(4^+)$ is
  negative and the cumulative effects of all terms are expected to
  make $\Delta e(4^+)$ large enough in absolute value to overcome $\Delta
  E_0 > 0$. This explains why the deformation energy curve of the $4^+$
  configuration has a local minimum at finite octupole
  deformation (scenario~(c)).

In contrast, the $K^{\pi} = 5^+$ $(3/2^+[651])_n
(7/2^+[404])_p$ configuration involves the creation of a neutron on
the $3/2^+[651]$ level above $N=88$ and the same proton excitation as
the $4^+$ configuration. This leads to the following total single-particle
energy correction
\begin{subequations}
  \begin{align}
    \label{eq_Deltae5+}
  \Delta e(5^+) & = \Delta e^{(\rm n)}_{3/2^+[651]} +
  \Delta e^{(\rm p)}_{7/2^+[404]}
  - 2\,\Delta e^{(\rm p)}_{5/2^-[523]} < 0 \\
  & = \Delta e(4^+) +
  \Delta e^{(\rm n)}_{1/2^+[640]} -
  \Delta e^{(\rm n)}_{3/2^+[651]} \,.
\end{align}
\end{subequations}
The quantity $\Delta e^{(\rm n)}_{1/2^+[640]} -
  \Delta e^{(\rm n)}_{3/2^+[651]}$ is fairly large as suggested by
the strongly increasing $1/2^+[640]$ neutron single-particle energy in
the Nilsson diagram of figure~\ref{fig_Gd154_Nilsson_diagrams}.
Therefore either $\Delta e(5^+)$, which is negative according to
Eq.~(\ref{eq_Deltae5+}), is expected to be rather small in absolute value and
unable to overcome $\Delta E_0$, so we end up with the scenario~(a).

\subsection{Octupole deformation in some odd-mass and odd-odd nuclei near $A=230$}

\label{subsec_actinides}

We present the results in this mass region in three parts: first, we
deal with the $N=139$ doubly-odd nuclei of our selection, then we discuss
the $7/2^-$ and $3/2^+$ neutron configurations in two odd-mass nuclei,
finally we focus on the proton parity doublet of $1/2$ states near
$Z=88$ in two odd-$Z$ nuclei.

\begin{figure}[b]
  \centering
  \includegraphics[width=\textwidth]{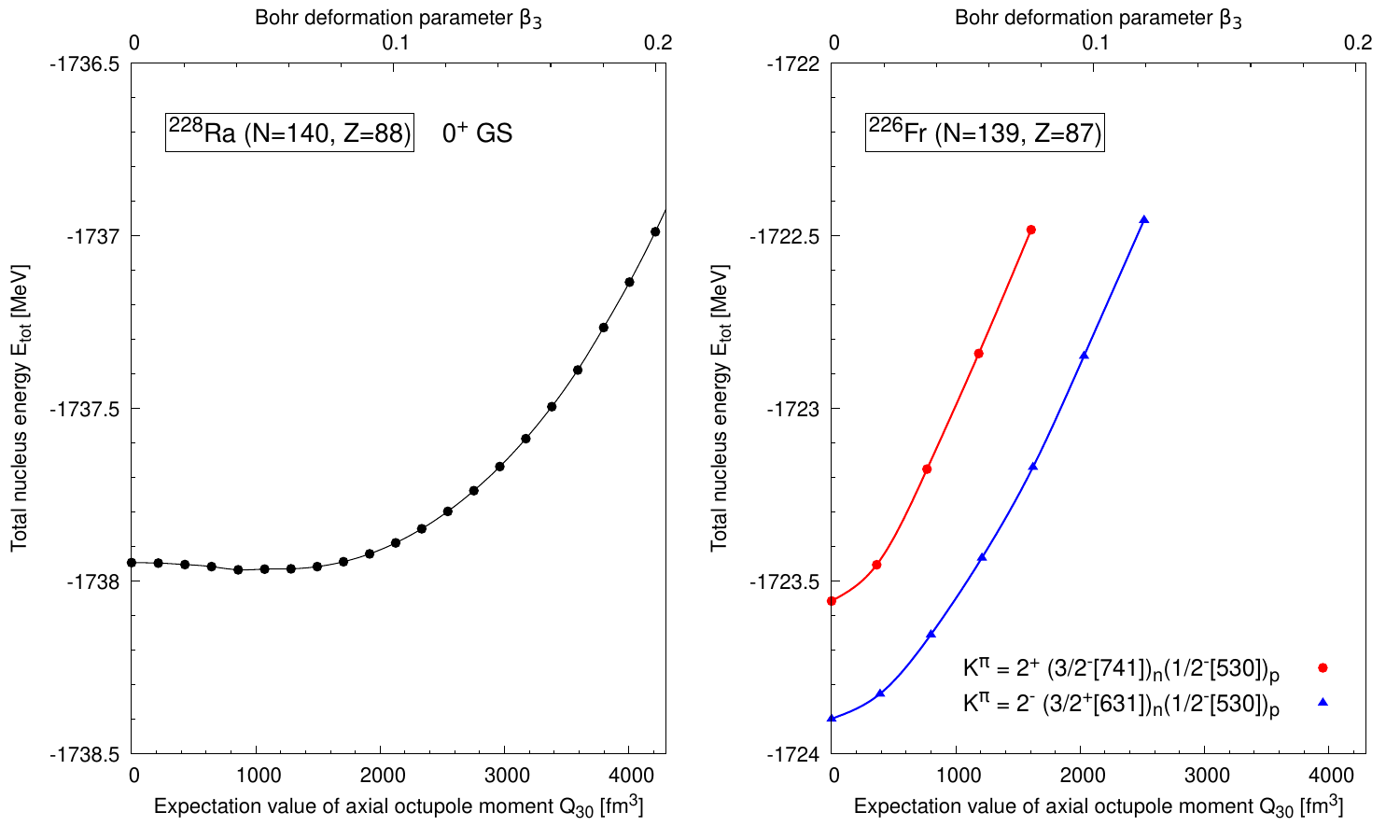}
\caption{Same as figure~\ref{fig_cutQ30_Ba144_Ba145} for the $^{226}$Fr
  nucleus and the reference nucleus $^{228}$Ra.
  \label{cutQ30_Ra228_Fr226}}
\end{figure}

\begin{figure}[ht]
  \centering
  \includegraphics[width=\textwidth]{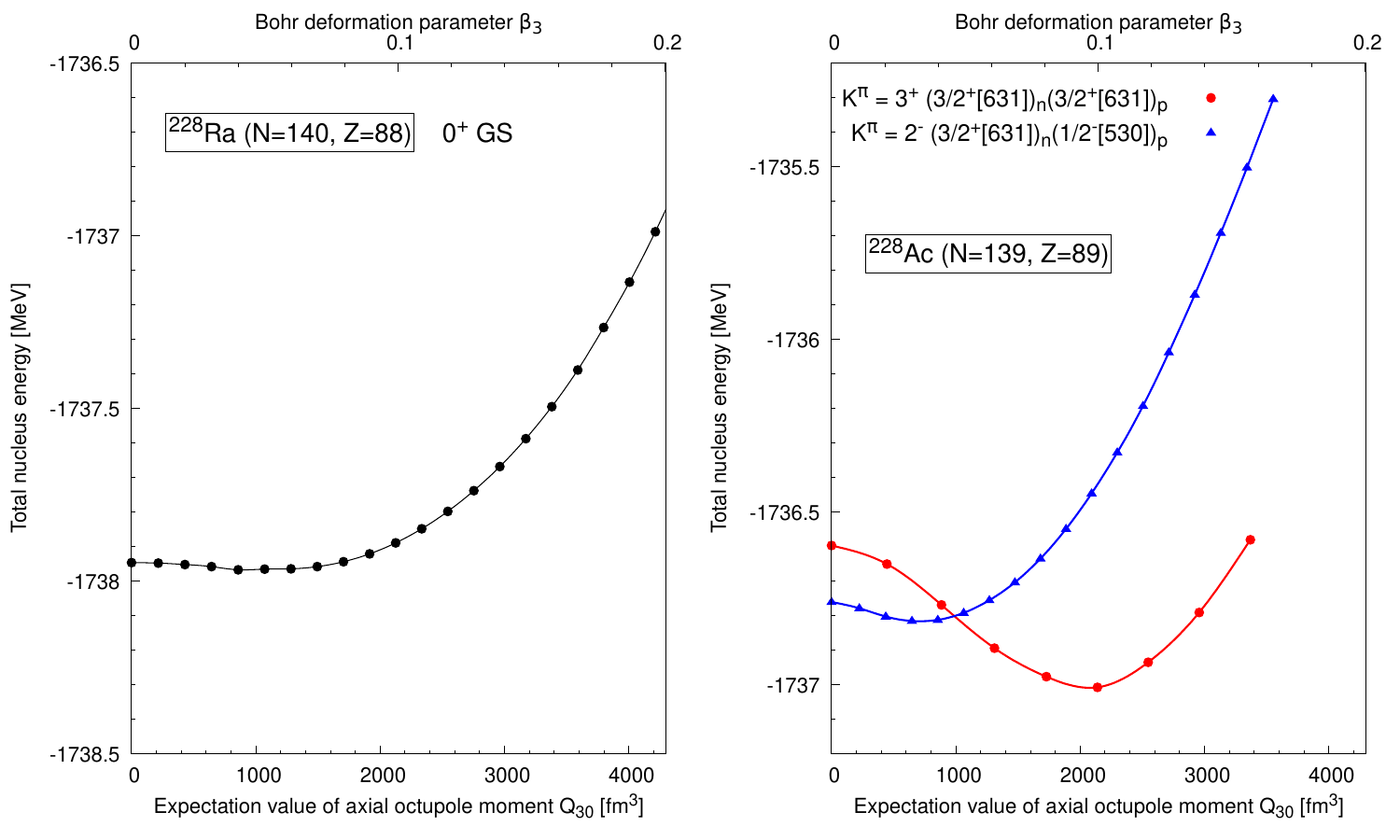}
\caption{Same as figure~\ref{fig_cutQ30_Ba144_Ba145} for the $^{228}$Ac
  nucleus and the reference nucleus $^{228}$Ra.
  \label{cutQ30_Ra228_Ac228}}
\end{figure}

\begin{figure}[h]
  \centering
  \includegraphics[width=\textwidth]{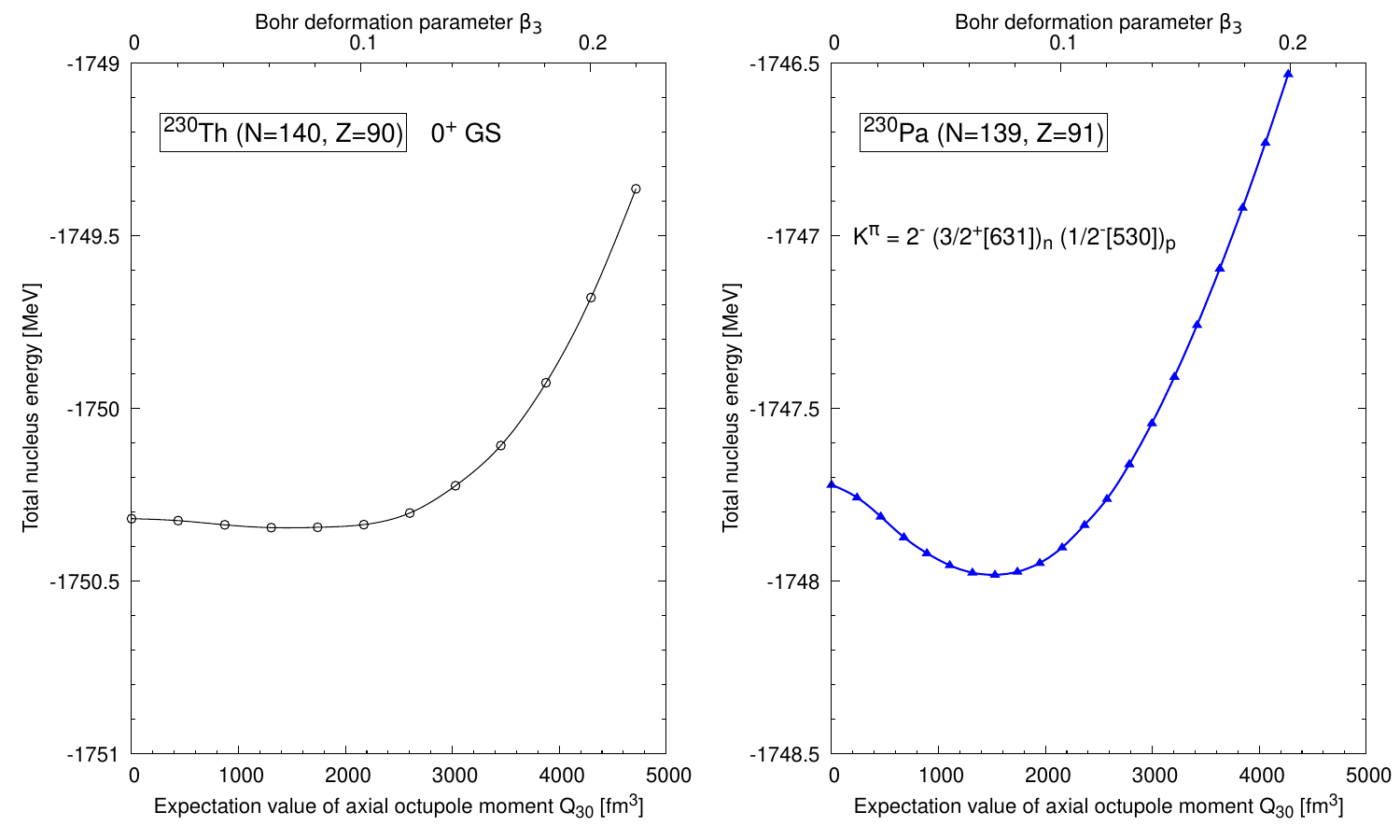}
\caption{Same as figure~\ref{fig_cutQ30_Ba144_Ba145} for the $^{230}$Pa
  nucleus and the reference nucleus $^{230}$Th.
  \label{cutQ30_Th230_Pa230}}
\end{figure}

\paragraph{Configurations in $N=139$ odd-odd isotones.}
First we discuss the $K^{\pi} = 2^-$
$(3/2^+[631])_n(1/2^-[530])_p$ configuration considered in three $N=139$ isotone
odd-odd nuclei, namely \ce{^{226}_{87}Fr_{139}},
\ce{^{228}_{89}Ac_{139}} and \ce{^{230}_{91}Pa_{139}}. As can seen in 
figures~\ref{cutQ30_Ra228_Fr226}, \ref{cutQ30_Ra228_Ac228} and
\ref{cutQ30_Th230_Pa230}, respectively, very different behaviors of
the total nucleus energy as a function of octupole deformation result
from the Hartree--Fock--BCS calculations.
To explain this feature, let us apply the perturbative mechanism using
the $^{228}$Ra even-even reference nucleus for $^{226}$Fr and
$^{228}$Ac and the $^{230}$Th reference nucleus for $^{230}$Pa.

On the one hand, according to the Nilsson diagrams of
figure~\ref{fig_Ra228_Nilsson_diagrams}, we can write for $^{226}$Fr
and $^{228}$Ac
\begin{align}
  \label{eq_2-_Fr226}
\Delta e(2^-,^{226}\!{\rm Fr}) & = - \Delta e^{(\rm n)}_{3/2^+[631]} +
\Delta e^{(\rm p)}_{1/2^-[530]} - 2\,\Delta e^{(\rm p)}_{1/2^+[640]} \\
\Delta e(2^-,^{228}\!{\rm Ac}) & = - \Delta e^{(\rm n)}_{3/2^+[631]}
+ \Delta e^{(\rm p)}_{1/2^-[530]} 
\end{align}
with $\Delta e^{(\rm n)}_{3/2^+[631]} \sim \Delta e^{(\rm
  p)}_{1/2^-[530]} \sim -\Delta e^{(\rm p)}_{1/2^+[640]} > 0$. 
On the other hand, according to the Nilsson diagrams of
figure~\ref{fig_Th230_Nilsson_diagrams}, we have for $^{230}$Pa
\eq{
  \Delta e(2^-,^{230}\!{\rm Pa}) = - \Delta e^{(\rm n)}_{3/2^+[631]} +
  2\,\Delta e_{3/2^+[631]}^{(\rm p)} - \Delta e^{(\rm p)}_{1/2^-[530]}
}
$\Delta e^{(\rm p)}_{3/2^+[631]} \lesssim 0$, therefore $\Delta
e(2^-,^{230}\!{\rm Pa}) < 0$ and is expected to be larger in absolute
value than $\Delta e(2^-,^{228}\!{\rm Ac})$. Moreover the $^{228}$Ra
and $^{230}$Th nuclei are found to be very soft, hence $\Delta E_0
\sim 0$ for both. We thus obtain
\begin{align}
  E_{2^-,\mbox{\scriptsize $^{226}$Fr}}[\Phi'] & = E_{2^-,\mbox{\scriptsize $^{226}$Fr}}[\Phi]
  + \Delta E_0 + \Delta e(2^-,^{226}\mathrm{Fr})
  \nonumber \\
  & \approx E_{2^-,^{226}{\rm Fr}}[\Phi]
  + 2 \, \Delta e^{(\rm p)}_{1/2^-[530]}  > E_{2^-,^{226}{\rm Fr}}[\Phi] \\
  E_{2^-,\mbox{\scriptsize $^{228}$Ac}}[\Phi'] & = E_{2^-,\mbox{\scriptsize $^{228}$Ac}}[\Phi]
  + \Delta E_0 + \Delta e(2^-,^{228}\mathrm{Ac})
  \approx E_{2^-,\mbox{\scriptsize $^{228}$Ac}}[\Phi]  \,.
\end{align}
We thus conclude that the octupole-deformation-energy curve of the
$2^-$ configuration in $^{226}$Fr rises from the left-right symmetric
solution (scenario (a)), whereas it is approximately constant in
$^{228}$Ac, which is very soft (situation between scenarios (b) and (c)
because $^{228}$Ra is very soft), and significantly decreases in
$^{230}$Pa (situation between scenarios (b) and (c)
because $^{230}$Th is very soft).

Next we investigate the $K^{\pi} = 2^+$
$(3/2^-[741])_n(1/2^-[530])_p$ configuration of the odd-odd $^{226}$Fr
nucleus. The proton configuration is the same as in the
previous configuration, whereas the neutron blocked state is a hole
with respect to the \ce{^{228}_{88}Ra_{140}} reference
nucleus. The total single-particle energy correction is therefore
\eq{
  \Delta e(2^+) = - \Delta e^{(\rm n)}_{3/2^-[741]} +
\Delta e^{(\rm p)}_{1/2^-[530]} - 2\,\Delta e^{(\rm p)}_{1/2^+[640]}
}
with $\Delta e^{(\rm n)}_{3/2^-[741]} \lesssim 0$, hence $\Delta
e(2^-) > 0$. When adding $\Delta E_0(^{228}{\rm Ra}) \sim 0$ we
obtain the limiting case between the scenarios~(a) and (d) because
$^{228}$Ra is very soft.

In the $K^{\pi} = 3^+$ $(3/2^+[631])_n(3/2^+[631])_p$ configuration of
the odd-odd $^{228}$Ac nucleus, the neutron blocked state is a hole
state and the proton blocked state is a particle state with respect to
the ground-state configuration of $^{228}$Ra. We thus we have, using
the same line of reasoning as for the $2^-$ configuration in this
nucleus
\eq{
\Delta e(3^+) = -\Delta e^{(\rm n)}_{3/2^+[631]} +
\Delta e^{(\rm p)}_{3/2^+[631]}
}
with $\Delta e^{(\rm p)}_{3/2^+[631]} < 0$, therefore $\Delta e(3^+) <
0$, hence
\eq{
E_{3^+}[\Phi'] = E_{3^+}[\Phi] + \Delta E_0 + \Delta e(3^+) < E_{3^+}[\Phi]\,.
}
We end up with the limiting case between secnarios~(a) or (d) owing to
the octupole softness of the reference nucleus.

\paragraph{$K^{\pi}=7/2^-$ and $K^{\pi}=3/2^+$ neutron
  configurations.} These two configurations are studied in the two
odd-mass nuclei $^{229}$Th and $^{235}$U. The $7/2^-$ neutron
single-particle level is of particle character in both nuclei,
whereas the $3/2^+$ level is of hole character (in both
nuclei). However the blocking of either single-particle state requires
a different multi-particle--multi-hole excitation in $^{229}$Th and
$^{235}$U. Indeed, the $7/2^-$ configuration in $^{235}$U requires to
simply create a neutron 1-particle excitation on top of the ground-state
configuration of the $^{234}$U reference even-even nucleus, whereas in
$^{229}$Th it requires a 1-particle--2-hole excitation with respect to
the $^{230}$Th reference nucleus. The opposite is true to build the
$3/2^+$ configuration in these odd-mass nuclei.

\begin{figure}[h]
  \centering
  \includegraphics[width=\textwidth]{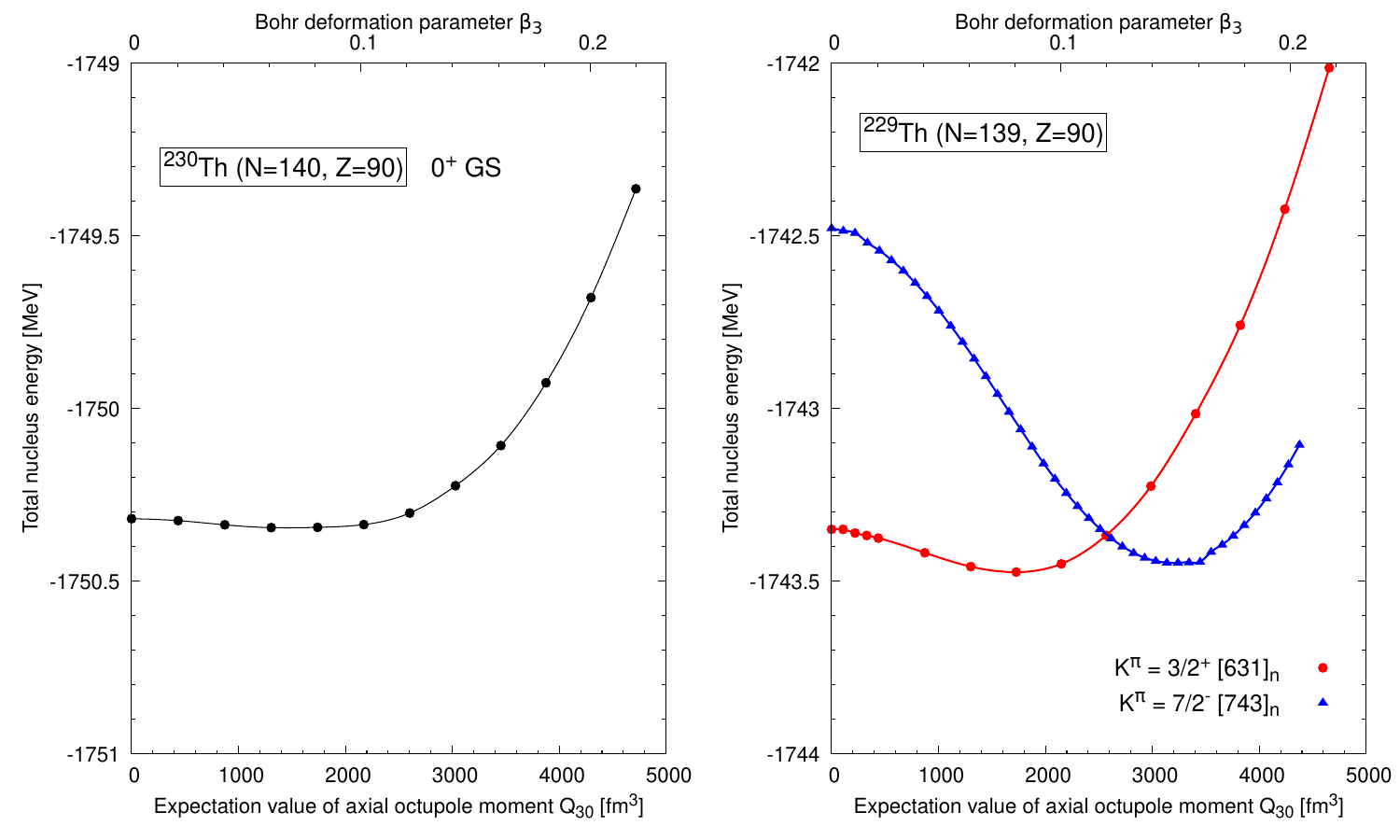}
\caption{Same as figure~\ref{fig_cutQ30_Ba144_Ba145} for the $^{229}$Th
  nucleus and the reference nucleus $^{230}$Th.
  \label{cutQ30_Th230_Th229}}
\end{figure}

The deformation energy curves obtained for the $3/2^+ [631]$ and
$7/2^-[743]$ neutron configurations in $^{229}$Th are plotted in
figure~\ref{cutQ30_Th230_Th229} as compared to the one of the
reference nucleus \ce{^{230}_{90}Th_{140}}.

Whereas a shallow minimum is found for the
former configuration, a very deep one appears for the 
latter. To understand these patterns in the framework of the
perturbative mechanism, we simply need to use the fact that in the
neutron spectrum of $^{230}$Th, the $7/2^-[743]$ level is of particle character
whereas the $3/2^+[631]$ level is of hole character. The $7/2^-$
configuration can thus be built from the ground-state configuration of
$^{230}$Th by annihilating the two $3/2^+$ last occupied neutrons and
creating a neutron in the $7/2^-$ level. The corresponding
single-particle correction energy is
\eq{
  \Delta e(7/2^-) = \Delta e_{7/2^-[743]}^{(\rm n)} - 2\,\Delta
  e_{3/2^+[631]}^{(\rm n)} \;.
}
In contrast, the $3/2^+$ configuration is simply described by a neutron
hole made in the $^{230}$Th nucleus, hence
\eq{
\Delta e(3/2^+) = - \Delta e_{3/2^+[631]}^{(\rm n)} \;.
}
If we make the approximation $\Delta e_{7/2^-[743]}^{(\rm n)} \approx -\Delta
e_{3/2^+[631]}^{(\rm n)} < 0$, we obtain
\eq{
  \Delta e(7/2^-) \approx - 3\,\Delta e_{3/2^+[631]}^{(\rm n)} \approx
  -3\,\Delta e(3/2^+) < 0\,.
}
Moreover $\Delta E_0 \approx 0$ for the very soft $^{230}$Th
nucleus. As a result, the deformation energy curves for both
configurations of $^{229}$Th have a local minimum at finite $Q_{30}$
and the one for the $7/2^-$ configuration should be much deeper than
the one for the $3/2^+$.   

The deformation energy curves obtained for the $3/2^+ [631]$ and
$7/2^-[743]$ neutron configurations in $^{235}$Th are plotted in
figure~\ref{cutQ30_Th230_Th229} as compared to the one of the
reference nucleus \ce{^{230}_{90}Th_{140}}.

\begin{figure}[ht]
  \centering
  \includegraphics[width=\textwidth]{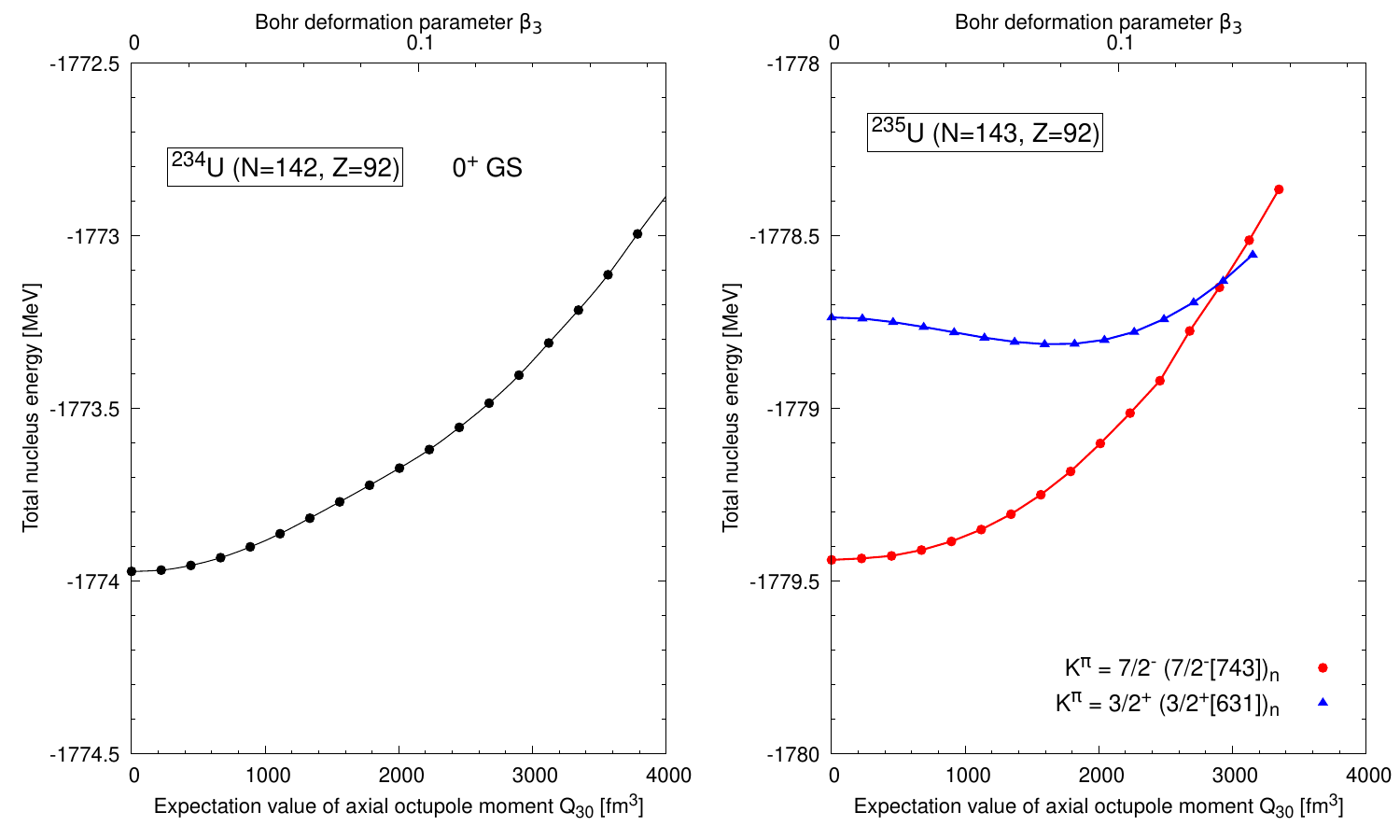}
\caption{Same as figure~\ref{fig_cutQ30_Ba144_Ba145} for the $^{235}$U
  nucleus and the reference nucleus $^{234}$U.
  \label{cutQ30_U234_U235}}
\end{figure}

\paragraph{Focus on the $1/2$ proton states near $Z=88$.}
We now investigate configurations involving the $1/2$ members of
the proton doublet near $Z=88$, namely the $1/2^+[640]$ and
$1/2^-[530]$ blocked state. We consider them in two nuclei, $^{231}$Ac
and $^{234}$Pa.

\begin{figure}[h]
  \centering
  \includegraphics[width=\textwidth]{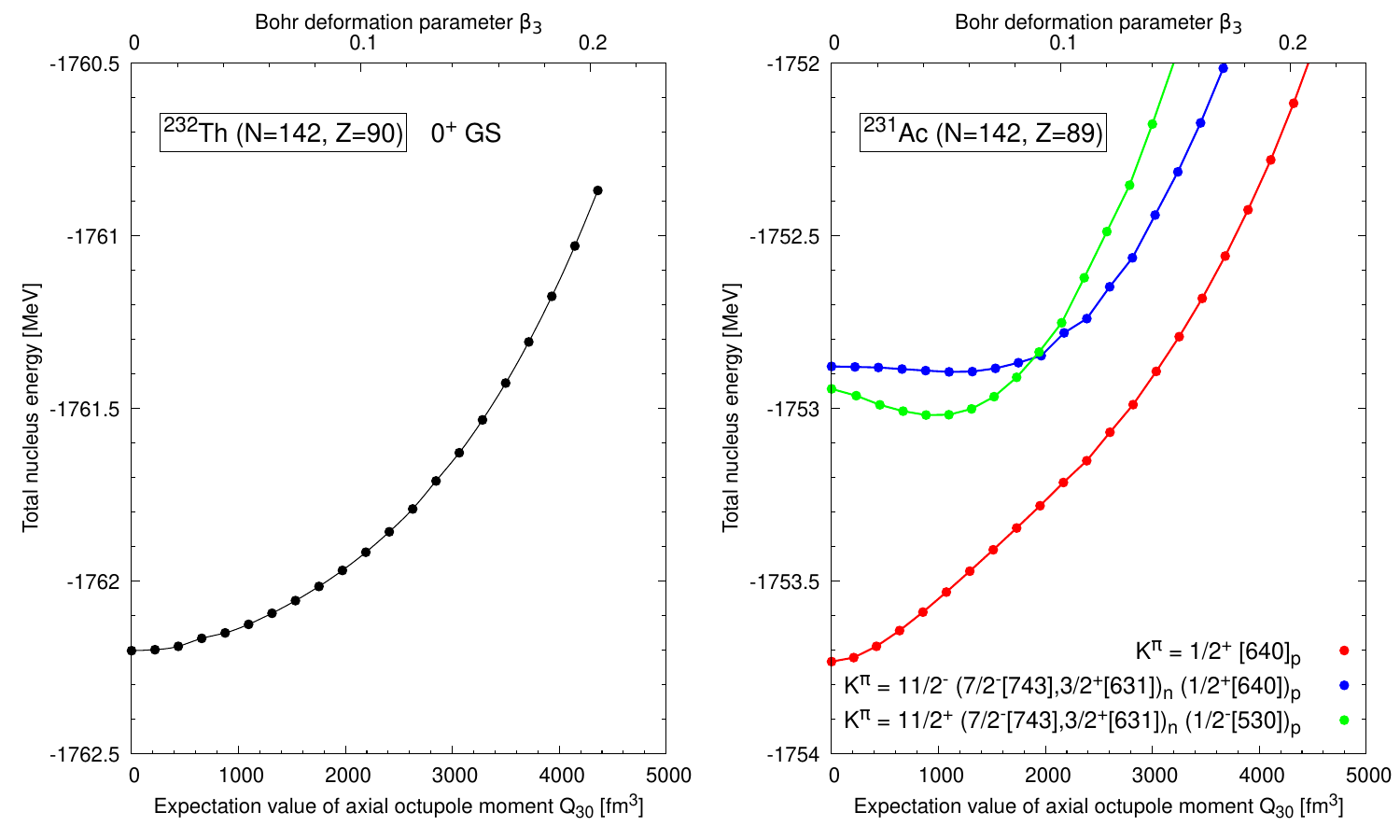}
\caption{Same as figure~\ref{fig_cutQ30_Ba144_Ba145} for the $^{231}$Ac
  nucleus and the reference nucleus $^{232}$Th.
  \label{cutQ30_Th232_Ac231}}
\end{figure}

First we focus on the $1/2^+[640]$ member of the proton parity doublet
$(1/2^+,1/2^-)$ near $Z = 88$ in $^{231}$Ac either alone as a
1-quasiparticle configuration (of hole character), or in addition to a
2-quasiparticle neutron configuration $K_n^{\pi_n} = 5^-$
$(7/2^-[743],3/2^+[631])$ to make a total $K^{\pi} = 11/2^-$
3-quasiparticle configuration. Then we compare with the
3-quasiparticle configuration obtained by replacing the $1/2^+[640]$
proton orbital by the $1/2^-[530]$ orbital. The resulting deformation
energy curves are displayed in figure~\ref{cutQ30_Th232_Ac231}. 

In the 1-quasiparticle configuration $K^{\pi} = 1/2^+$, the blocked
state $1/2^+[640]$ is a down-sloping orbital in the Nilsson diagram of
$^{232}$Th (see figure~\ref{fig_Th232_Nilsson_diagrams}), of hole
character. The single-particle energy correction is thus positive
\eq{
  \Delta e(1/2^+) = -\Delta e^{(\rm p)}_{1/2^+[640]}
}
since $\Delta e^{(\rm p)}_{1/2^+[640]} < 0$. Together with $\Delta
E_0$ in $^{232}$Th according to figure~\ref{cutQ30_Th232_Ac231}, the
perturbative mechanism therefore gives a deformation energy curve for
the $K^{\pi} = 1/2^+$ of $^{231}$Ac with a minimum at $Q_{30} = 0$
(scenario (a)) and stiffer than the one of the ground-state
configuration of $^{232}$Th, in agreement with actual
Hartree--Fock--BCS calculations.

With the addition of the neutron 2-quasiparticle $5^-$ configuration
of 1-particle--1-hole character $(7/2^-[743],3/2^+[631])$, we bring a
sizeably negative contribution to the single-particle energy
correction in the resulting $K^{\pi} = 11/2^-$ 3-quasiparticle
configuration
\eq{
\Delta e(11/2^-) = \Delta e^{(\rm n)}_{7/2^-[743]} -
\Delta e^{(\rm n)}_{3/2^+[631]} - \Delta e^{(\rm p)}_{1/2^+[640]} \,,
}
which compensates at least partly $\Delta E_0$. The resulting
deformation energy curve is expected to be rather flat at small
octupole deformation. This is precisely the pattern found in
figure~\ref{cutQ30_Th232_Ac231} corresponding to the limiting case
between scenarios (a) and (c).

To counterbalance the positive contribution of $\Delta E_0$, an extra
negative single-particle contribution can be brought in $^{231}$Ac by
replacing the $1/2^+[640]$ proton orbital with the $1/2^-[530]$
state. This does produce the small dip in the deformation energy
curve for the resulting $K^{\pi} = 11/^+$ configuration observed in
figure~\ref{cutQ30_Th232_Ac231} corresponding to the scenario (c).

\begin{figure}[h]
  \centering
  \includegraphics[width=\textwidth]{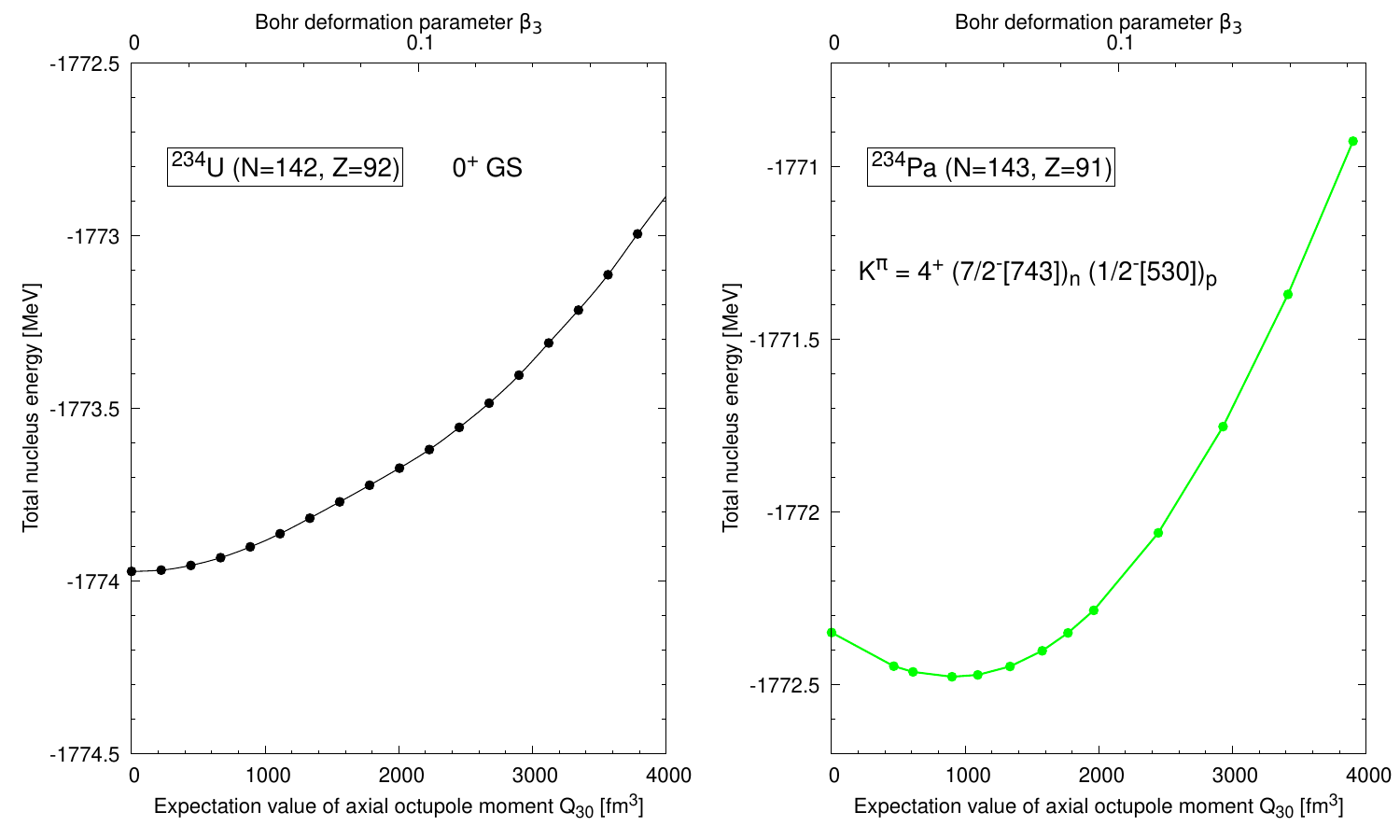}
\caption{Same as figure~\ref{fig_cutQ30_Ba144_Ba145} for the $^{234}$Pa
  nucleus and the reference nucleus $^{234}$U.
  \label{cutQ30_U234_Pa234}}
\end{figure}

Finally we turn to the $^{234}$Pa nucleus with the $K^{\pi} = 4^+$
$(7/2^-[743])_n(1/2^-[530])_p$ configuration, which is interpreted as
a $\rm (1p0h)_n(0p1h)_n$ excitation with respect to the ground-state
configuration of the $^{34}$U reference even-even nucleus. The
resulting deformation energy curve is plotted in
figure~\ref{cutQ30_U234_Pa234}. 
The perturbative mechanism interpretes this configuration as the
creation of a neutron in the $7/2^-[743]$ state of particle character
and the annihilation of a proton in the $1/2^-[530]$ state of hole
character. This gives
\eq{
  \Delta e = \Delta e^{(\rm n)}_{7/2^-[743]} - \Delta e^{(\rm p)}_{1/2^-[530]} \,,
}
with $\Delta e^{(\rm n)}_{7/2^-[743]} \sim - \Delta e^{(\rm
  p)}_{1/2^-[530]} < 0$ according to the Nilsson diagrams of the
$^{234}$U reference even-even nucleus in
figure~\ref{fig_U234_Nilsson_diagrams}. Moreover the $^{234}$U is
found to be less stiff against octupole deformation than $^{232}$Th as
can be seen from a smaller curvature of the $^{234}$U nucleus around
$\beta_3 =0$ than the curvature found in $^{232}$Th. We thus have
\eq{
  \Delta E_0(^{232}\mathrm{Th}) > \Delta E_0(^{232}\mathrm{Th}) > 0
}
The net effect $\Delta E_0+\Delta e$ in $^{235}$U is thus expected to
be in favor of the single-particle contribution, which leads to a
local minimum in the deformation energy curve at finite $\beta_3$
(scenario (c)).

\subsection{Discussion}

\begin{figure}[h]
  \centering
  \includegraphics[width=\textwidth]{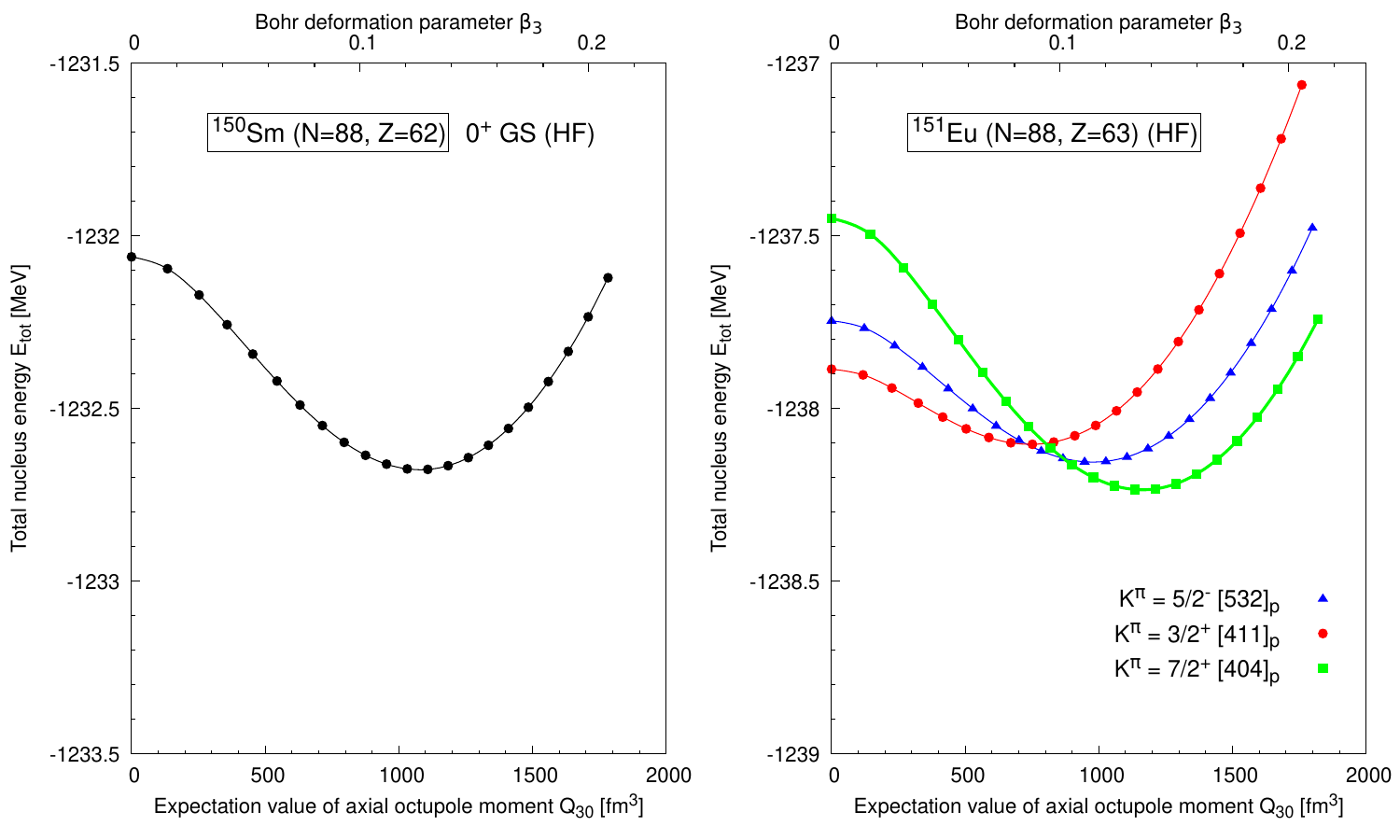}
\caption{Same as figure~\ref{fig_cutQ30_Sm150_Eu151} within the
  Hartree--Fock approach only (no pairing).
  \label{fig_cutQ30_Sm150_Eu151_HF}}
\end{figure}

To test the Koopmans approximation which is at the heart of the
perturbative mechanism, we have repeated the calculation of the
octupole-deformation-energy curve without pairing, that is within the
Hartree--Fock approximation. We have chosen a nucleus for which the
convergence of selfconsistent calculations with blocking and parity
breaking is possible, namely $^{151}$Eu. The resulting curve is
plotted in figure~\ref{fig_cutQ30_Sm150_Eu151_HF}, together with the
deformation energy curve of the $^{150}$Sm reference even-even nucleus.

It can be noticed that the three considered configurations have a
local minimum at finite octupole deformation and that the observed
relative depths of these minima confirm the expectations based on the
$\Delta e^{(\rm n)}$ values of
table~\ref{tab_Q30_sensitive_orbitals}. This was not the case within
Hartree--Fock--BCS calculations, therefore pairing plays a non
negligible role in this nucleus. This calls for an improvement of the
deformation mechanism presented here. 

\begin{figure}[ht]
  \includegraphics[width=0.5\textwidth]{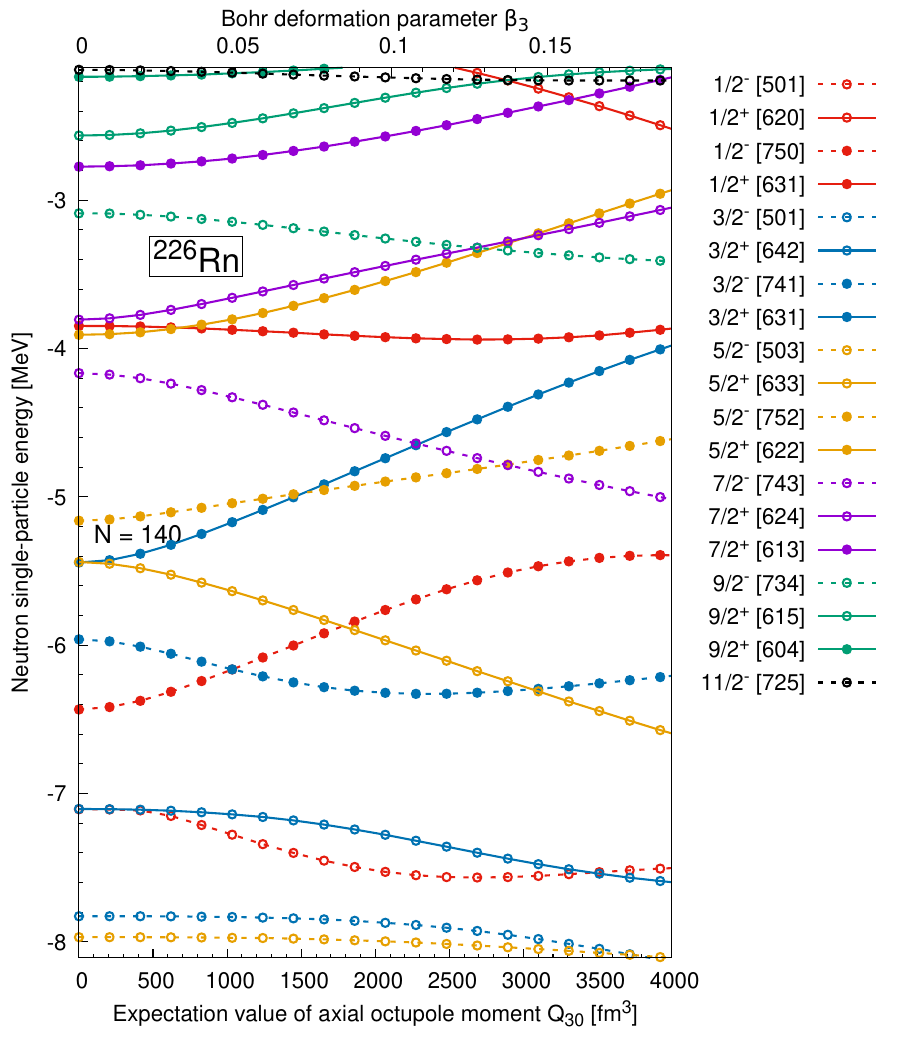} 
  \includegraphics[width=0.5\textwidth]{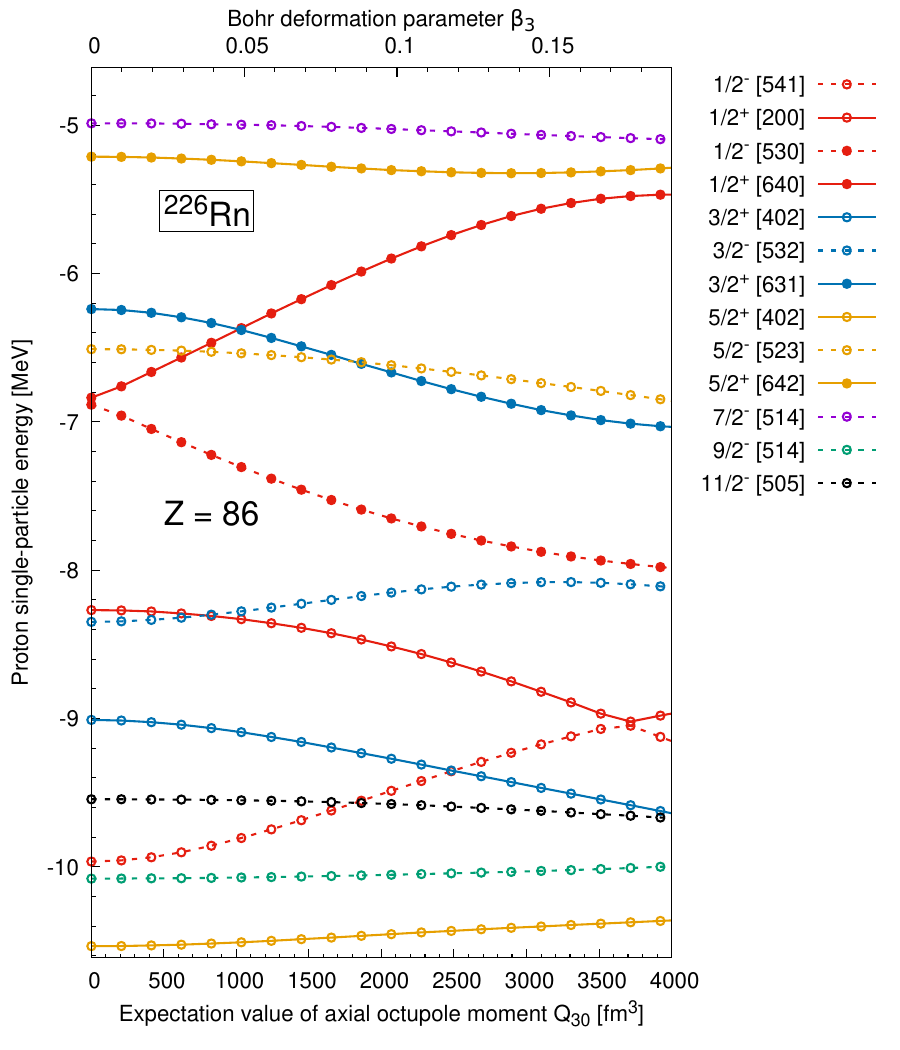}
  \caption{\label{fig_Rn226_Nilsson_diagrams} Same as
    figure~\ref{fig_Ba144_Nilsson_diagrams} for the
    \ce{^{226}_{86}Rn_{140}} nucleus.}
\end{figure}

Another important aspect of the perturbative mechanism is the choice
of the reference even-even nucleus. In this respect, a peculiar
feature is the inversion of the two proton $1/2$ states across the
$Z=88$ shell gap in $^{226}$Rn and the other even-even nuclei with $Z
> 86$, as can be seen in figure~\ref{fig_Rn226_Nilsson_diagrams}.
In the former nucleus the $1/2^-[530]$ orbital lies below the
$1/2^+[640]$ in the single-particle spectrum and these two levels are
almost degenerate, whereas in the heavier nuclei this is the other way
around and the energy difference between them is larger as $Z$
increases. This may lead to an ambiguity of interpretation of proton
configurations involving either member of this $1/2$ proton 
parity doublet.

Let us try to apply the perturbative mechanism using the $^{226}$Rn and
$^{230}$Th reference nuclei to try to explain this feature.

On the one hand, according to the Nilsson
diagrams of $^{226}$Rn and $^{230}$Th, we can write
\begin{align}
  \label{eq_2-_Fr226}
\Delta e(2^-,^{226}\!{\rm Fr}) & = \Delta e^{(\rm p)}_{1/2^-[530]}(^{226}{\rm Rn}) -
\Delta e^{(\rm n)}_{3/2^+[631]}(^{226}{\rm Rn}) \\
\Delta e(2^-,^{228}\!{\rm Ac}) & = - \Delta e^{(\rm p)}_{1/2^-[530]}(^{230}{\rm Th}) -
\Delta e^{(\rm n)}_{3/2^+[631]}(^{230}{\rm Th})
\lesssim -2\,\Delta e^{(\rm p)}_{3/2^+[631]}(^{230}{\rm Th})
< 0
\end{align}
using the inequality $\Delta e^{(\rm p)}_{1/2^-[530]} \gtrsim \Delta e^{(\rm
  n)}_{3/2^+[631]}$ as suggested by the Nilsson diagrams of
$^{230}$Th.
However, as noticed at the end of subsection
\ref{subsec_e-e_Q30_curves}, the $1/2^-[530]$ proton level in
$^{226}$Rn is almost degenerate with the $1/2^+[640]$ level and is
just below the latter, in contrast to the ordering found in
$^{230}$Th. Therefore one approximately has
\eq{
  \Delta e^{(\rm p)}_{1/2^-[530]}(^{226}{\rm Rn}) \sim
  - \Delta e^{(\rm p)}_{1/2^-[530]}(^{230}{\rm Th}) < 0 \,.
}
There is thus an ambiguity of application of the perturbative
mechanism. As a matter of fact, the $1/2^-[530]$ proton blocked state
lies above the $1/2^+[640]$ state in the left-right symmetric
Hartree--Fock--BCS solution of $^{226}$Fr with selfconsistent blocking
for the $K^{\pi} = 2^-$ $(3/2^+[631])_n(1/2^-[530])_p$
configuration. Therefore it is more physical to consider that $\Delta e^{(\rm
  p)}_{1/2^-[530]}$ is positive than $\Delta e^{(\rm
  p)}_{1/2^-[530]} < 0$. Let us thus continue the analysis with $\Delta e^{(\rm
  p)}_{1/2^-[530]}(^{226}{\rm Rn}) \sim \Delta e^{(\rm
  p)}_{1/2^-[530]}(^{230}{\rm Th}) > 0$.
Using again the inequality $\big|\Delta e^{(\rm p)}_{1/2^-[530]}\big|
\gtrsim \Delta e^{(\rm n)}_{3/2^+[631]} > 0$, Eq.~(\ref{eq_2-_Fr226})
thus becomes 
\eq{
\Delta e(2^-,^{226}\!{\rm Fr}) \gtrsim 0 \,.
}

\begin{figure}[ht]
  \centering
  \includegraphics[width=0.6\textwidth]{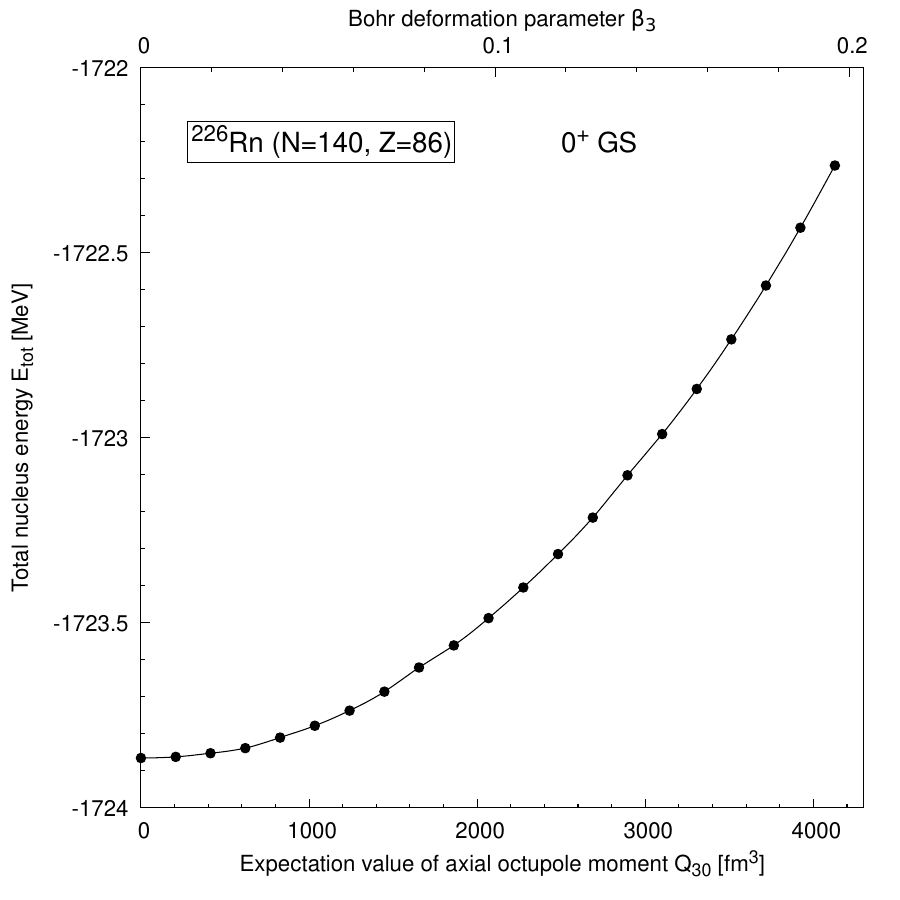} 
  \caption{\label{cutQ30_Rn226} Deformation energy curve of the
    \ce{^{226}_{86}Rn_{140}} nucleus in its ground-state configuration
    calculated within the Hartree--Fock--BCS approach.}
\end{figure}

On the other hand, octupole deformation is not
energetically favorable in the ground-state configuration of
$^{226}$Rn according to figure~\ref{cutQ30_Rn226}, hence
$\Delta E_0(^{226}\rm Rn) > 0$, and the $^{230}$Th nucleus is very
soft, hence $\Delta E_0(^{230}\rm Th) \sim 0$. We thus obtain
\begin{align}
  E_{2^-,\mbox{\scriptsize $^{226}$Fr}}[\Phi'] & = E_{2^-,\mbox{\scriptsize $^{226}$Fr}}[\Phi]
  + \Delta E_0(^{226}\rm Rn) + \Delta e(2^-,^{226}\mathrm{Fr})
  \nonumber \\
  & \approx E_{2^-,^{226}{\rm Fr}}[\Phi]
  + \Delta E_0(^{226}{\rm Rn}) > E_{2^-,^{226}{\rm Fr}}[\Phi] \\
  E_{2^-,\mbox{\scriptsize $^{228}$Ac}}[\Phi'] & = E_{2^-,\mbox{\scriptsize $^{228}$Ac}}[\Phi]
  + \Delta E_0(^{230}\rm Th) + \Delta e(2^-,^{228}\mathrm{Ac})
  \nonumber \\
  & \approx E_{2^-,\mbox{\scriptsize $^{228}$Ac}}[\Phi] - 2 \, \Delta e^{(\rm
    p)}_{3/2^+[631]} < E_{2^-,\mbox{\scriptsize $^{228}$Ac}}[\Phi] \,.
\end{align}

We thus conclude that the octupole-deformation-energy curve of the
$2^-$ configuration in $^{226}$Fr rises from the left-right symmetric
solution (scenario (a)) whereas it decreases in $^{228}$Ac (situation
between scenarios (b) and (c) because $^{230}$Th is very soft).

%
%
%
%

\section{Conclusion and perspectives}

In this work the mechanism proposed in
Ref.~\cite{Bonneau24_PhysScr99} to explain the presence or absence of
octupole deformation in 2-quasiparticle states of even-even nuclei is
generalized to multi-quasiparticle configurations of odd-mass and
doubly-odd nuclei. It relies on two main approximations. On the one
hand, we assume the Koopmans approximation, in other words we neglect
the role of pairing correlations. On the other hand we approximate the
single-particle Hamiltonian of selfconsistent octupole-constrained
Hartree--Fock--BCS solutions by the sum of the mean-field Hamiltonian
of the parity-conserved solution and a perturbing potential
proportional to the axial octupole moment operator $\widehat
Q_{30}$ (hence proportional to the Bohr octupole deformation parameter
$\beta_3$). These assumptions allow to express the variation with
$Q_{30}$ of the total energy of a nucleus in a given configuration
$\mathcal C$ for small deformations as the sum of the corresponding
variation $\Delta E_0$ in a reference even-even nucleus and
the total single-particle energy correction $\Delta e$ at second order
in $\beta_3$ involving all the states of the
multi-particle--multi-hole excitation describing $\mathcal C$ with
respect to the reference nucleus.

Through the application of the mechanism to two mass regions known to
exhibit octupole deformation in even-even nuclei with neutron numbers
near $N = 88$, 138 and proton numbers near $Z = 56$, 88, we show that
the perturbative mechanism is able to explain qualitatively features
observed in the octupole-deformation-energy curves of most odd-mass
and odd-odd nuclei. The balance between $\Delta E_0$ and $\Delta e$,
when their signs differ, can sometimes be left undecided depending on
the sensitivity of the involved single-particle states to octupole
perturbation, but in many cases the estimates of these two quantities
provide a consistent explanation of the variety of behaviors in the
constrained Hartree--Fock--BCS calculations.

Despite this overall success, we have discussed two main limitations in the
peturbative mechanism through two examples. First, the pairing
correlations play a role that remains to be incorporated. This has been
shown by repeating the analysis for one odd-mass nucleus within the
Hartree--Fock framework (no pairing), with which the predictions of
the perturbative mechanism compares more favorably than with the
Hartree--Fock--BCS results. Second, the choice of the
reference even-even nucleus to interprete the blocked configuration as
a multi-particle--multi-hole excitation can sometimes influence the
prediction of the mechanism. This happens when the blocked states are
almost degenerate with other single-particle states. In this case, the
implementation of the perturbative mechanism requires a reference
nucleus in which the shell gap at Fermi level, away from the blocked
states, is large enough to avoid the ambiguity of the particle or hole
character of blocked states, which can somehow artificially increase
the multi-particle--multi-hole excitation order of the considered
configuration.

Finally, in some configurations, calculations with selfconsistent blocking
and parity breaking do not converge because of the strong $\widehat
Q_{30}$ coupling between two single-particle states at small octupole
deformation. This coupling induces two different problems. On the one
hand, this produces a large parity mixing of different amplitudes in
the $\Omega$ and $-\Omega$ states, making the identification of
canonical conjugate states ambiguous when using the criterion of
maximal overlap of the $\Omega$ state with the time-reversal of the
candidate $-\Omega$ state. On the other hand, when the average parity  
reaches 0 in two octupole-coupled states (very large parity mixing),
the identification of the
state to be blocked becomes very difficult, even using the continuity
of its single-particle energy as a function of $\beta_3$. The
configurations leading to these problems have been discared in the
present work and deserve special attention in a future work.

\appendix

\section*{Acknowledgements}
We warmly thank P. Quentin for fruitful discussions. 
Support by the Bulgarian National Science Fund (BNSF) under Contract No. KP-06-N98/2 is acknowledged.

\section{Axially-deformed harmonic oscillator basis}

The single-nucleon states of isospin index $\tau = n$ for neutrons or
$\tau = p$ for protons, denoted by $\ket{\psi_x^{(\tau)}}$, are
expanded in the axially-deformed harmonic-oscillator
basis~\cite{FQKV} as
\eq{
  \label{psi_expansion_HO_basis}
  \ket{\psi_x^{(\tau)}} = \sum_{N,n_z,\Lambda,\Sigma}
  C_{Nn_z\Lambda\Sigma}^{(x,\tau)} \ket{Nn_z\Lambda\Sigma}
}
where $n_z$ is the number of oscillator quanta in the $z$ direction
chosen as the symmetry axis in the intrinsic frame, $N = n_z +
n_{\bot}$ with $n_{\bot}$ the number of oscillator quanta in the
direction orthogonal to the symmetry axis, $\Lambda$ is the eigenvalue
in $\hbar$ unit of the $z$ component of the orbital angular momentum
$\widehat L_z$ and $\Sigma = \pm 1/2$ is the eigenvalue in $\hbar$
unit of the $z$ component of the spin angular momentum $\widehat
S_z$. It is worth noting that basis states $\ket{N n_z \Lambda
  \Sigma}$ are such that $|\Lambda| \leqslant n_{\bot}$.

Because of invariance by rotation around the axis $z$, the sum
$\Lambda + \Sigma = \Omega$, which represents the eigenvalue in
$\hbar$ unit of the total angular momentum $\widehat J_z = \widehat
L_z + \widehat S_z$, is a good quantum number. We thus have 
\eq{
  \widehat J_z \ket{\psi_x^{(\tau)}} = \Omega_x \hbar \ket{\psi_x^{(\tau)}}
}
and the expansion (\ref{psi_expansion_HO_basis}) of
$\ket{\psi_x^{(\tau)}}$ is only a triple sum
\eq{
  \label{psi_expansion_HO_basis_axial}
  \ket{\psi_x^{(\tau)}} = \sum_{N,n_z,\Lambda}\sum_{\Sigma=\pm 1/2}
  C_{Nn_z\Lambda\Sigma}^{(x,\tau)} \ket{Nn_z\Lambda\Sigma} \qquad \mbox{with
    $\Lambda = \Omega_x - \Sigma$}
}

The spatial part $\ket{N n_z \Lambda}$ of the basis state
$\ket{Nn_z\Lambda\Sigma}$ is characterized by the oscillator
frequency $\omega_z$ along the symmetry axis and the frequency
$\omega_{\bot}$ in the direction perpendicular to the symmetry
axis. As done in Ref.~\cite{FQKV}, we instead work with the
basis-deformation parameter
\eq{
  q = \frac{\omega_{\bot}}{\omega_z}
}
and the inverse $b$ of the oscillator length associated with the
frequency $\omega_0 = (\omega_z\omega_{\bot}^2)^{1/3}$, namely
\eq{
  b = \sqrt{\frac{m\omega_0}{\hbar}}
}
with $m$ the nucleon mass.

In practice the basis has to be
truncated. To do so we use a truncation scheme in which the sums over
$N$ and $n_z$ in the expansion (\ref{psi_expansion_HO_basis_axial})
run from 0 to cutoff values deduced from the following
relation~\cite{FQKV,Damgaard69}
\eq{
  \hbar\omega_z(n_z + 1/2) + \hbar\omega_{\bot}(n_{\bot} + 1)
  \leqslant \hbar\omega_0(N_0+2)
}
where $N_0+1$ represents the number of major oscillator shells. The
basis parameters $b$ and~$q$ are optimized in the considered
even-even reference nuclei by minimization of their total energy
in the ground-state configuration.

\section{Matrix elements of $\widehat Q_{30}$ between axially-deformed
  harmonic oscillator states}

Let us first introduce the oscillator parameters along the symmetry
axis $z$
\eq{
  \beta_z = \sqrt{\frac{m\omega_z}{\hbar}} = b\,q^{-1/3}
}
and in the perpendicular plane $xy$
\eq{
  \beta_{\perp} = \sqrt{\frac{m\omega_{\perp}}{\hbar}} = b\,q^{1/6}
  \,. 
}
Following the method of Quentin and Babinet~\cite{Quentin-Babinet70},
using a bosonic representation of operators in the axially-deformed
harmonic oscillator basis, we can express the axial-octupole moment
operator $\widehat Q_{30}$ defined in Eq.~(\ref{eq_Q30_def}) as
\begin{align}
  \widehat Q_{30} = \sqrt{\frac{7}{16\pi}}\,
  \frac{1}{\beta_z\sqrt{2}}\Big\{ &
  \frac{1}{\beta_z^2}\Big[3\big(a^\dagger_z(\widehat
    N_z + 1) + a_z\widehat N_z\big) + {a_z^\dagger}^3 + a_z^3\Big]
  \nonumber \\
  & -\frac{3}{\beta_\perp^2}(a^\dagger_z + a_z)\big(1 + \widehat
    N_\perp + \widehat N_+ + \widehat N_-\big)\Big\}\,,
\end{align}
where the boson operator $a^{\dagger}_z$ (respectively $a_z$) creates 
(respectively annihilates) an oscillator quantum in the $z$ direction,
the operator $\widehat N_z = a^{\dagger}_za_z$ counts the number of
oscillator quanta in the $z$ direction, the operator $\widehat
N_{\perp}$ counts the number of oscillator quanta in the $xy$ plane, and
$\widehat N_{\pm}$ behave as double-ladder operators in the $xy$
plane. These operators are defined by their action on a basis
state $\ket{N n_z \Lambda \Sigma}$ according to
\begin{align}
  a^{\dagger}_z\ket{Nn_z\Lambda\Sigma} & = \sqrt{n_z+1} \,
  \ket{Nn_z\Lambda\Sigma} \\
  a_z\ket{Nn_z\Lambda\Sigma} & = \sqrt{n_z} \, 
  \ket{Nn_z\Lambda\Sigma} \\
  \widehat N_z\ket{Nn_z\Lambda\Sigma} & = n_z \,
  \ket{Nn_z\Lambda\Sigma} \\
  \widehat N_{\perp}\ket{Nn_z\Lambda\Sigma} & = n_{\perp} \, 
  \ket{Nn_z\Lambda\Sigma} \\
  \widehat N_+\,\ket{Nn_z\Lambda\Sigma} & = -\dfrac 1 2
  \sqrt{(n_{\perp}+2)^2 - \Lambda^2} \,
  \ket{N\!+\!2 \,n_z\Lambda\Sigma} \\
  \widehat N_-\,\ket{Nn_z\Lambda\Sigma} & = -\dfrac 1 2
  \sqrt{n_{\perp}^2 - \Lambda^2} \,
  \ket{N\!-\!2 \,n_z\Lambda\Sigma}\,.
\end{align}
with $n_{\perp} = N - n_z$. It is recalled that $|\Lambda| \leqslant
n_{\bot}$. The matrix element of $\widehat Q_{30}$ between the basis
states $\ket{Nn_z\Lambda\Sigma}$ and $\ket{Nn_z\Lambda\Sigma}$ is thus
given by
\eq{
  \elmx{N'\,n_z'\,\Lambda'\,\Sigma'}{\widehat
    Q_{30}}{N\,n_z\,\Lambda\,\Sigma} =
  \delta_{\Lambda\Lambda'}\,\delta_{\Sigma\Sigma'} \,
  \elmx{N'\,n_z'\,\Lambda}{\widehat
    Q_{30}}{N\,n_z\,\Lambda}
}
where
\begin{align}
  \elmx{N'\,n_z'\,\Lambda}{\widehat Q_{30}}{N\,n_z\,\Lambda} =
  & \sqrt{\frac{7}{16\pi}}\,\frac{1}{\beta_z\sqrt{2}}\times \nonumber \\
  & \Big\{
  \frac{1}{\beta_z^2} \Big[
    3(n_z+1)^{\frac 32}\delta_{N'N+1}\,\delta_{n_z'n_z+1}
    + 3n_z^{\frac 3 2}\delta_{N'N-1}\,\delta_{n_z'n_z-1}  \nonumber \\
    & \phantom{\frac{1}{\beta_z^2} \Big[}
      + \sqrt{(n_z+1)(n_z+2)(n_z+3)}\,\delta_{N'N+3}\,\delta_{n_z'n_z+3}
      \nonumber \\ 
    & \phantom{\frac{1}{\beta_z^2} \Big[}
        + \sqrt{n_z(n_z-1)(n_z-2)}\,\delta_{N'N-3}\,\delta_{n_z'n_z-3}\Big]
      \nonumber \\
    & - \frac{3}{\beta_\perp^2}\Big[\sqrt{n_z + 1}
    \Big((n_\perp+1)\delta_{N'N+1}\,\delta_{n_z'n_z+1}
    \nonumber \\ 
    & \phantom{- \frac{3}{\beta_\perp^2}\Big[\sqrt{n_z + 1}
    \Big(}
    - \frac 1 2\,\sqrt{(n_{\perp}+2)^2-\Lambda^2} \,
    \delta_{N'N+3}\,\delta_{n_z'n_z+1} \nonumber \\
    & \phantom{- \frac{3}{\beta_\perp^2}\Big[\sqrt{n_z + 1}
    \Big(}
    - \frac 1 2\, \sqrt{n_{\perp}^2 - \Lambda^2} \,
    \delta_{N'N-1}\,\delta_{n_z'n_z + 1}\Big)  \nonumber \\
    & \phantom{- \frac{3}{\beta_\perp^2}\Big[}
      + \sqrt{n_z}\Big((n_\perp-1)\delta_{N'N-1}\,\delta_{n_z'n_z-1}
      \nonumber \\ 
    & \phantom{- \frac{3}{\beta_\perp^2}\Big[+ \sqrt{n_z}\Big(}
      - \frac 1 2\,\sqrt{(n_{\perp}+2)^2-\Lambda^2} \,
      \delta_{N'N+1}\,\delta_{n_z'n_z-1} \nonumber \\
    & \phantom{- \frac{3}{\beta_\perp^2}\Big[+ \sqrt{n_z}\Big(}
      - \frac 1 2\, \sqrt{n_{\perp}^2 - \Lambda^2} \,
      \delta_{N'N-3}\,\delta_{n_z'n_z - 1}\Big)\Big]\Big\}\,. 
\end{align}


\end{document}